\documentclass[amsmath,amssymb,aps,prx, twocolumn,superscriptaddress]{revtex4-2}

\RequirePackage{epsfig}
\RequirePackage{graphicx}

\RequirePackage{xcolor}
\RequirePackage{siunitx}
\RequirePackage{nicefrac}

\RequirePackage{booktabs}

\RequirePackage{enumerate}
\RequirePackage{enumitem}

\RequirePackage[normalem]{ulem}
\RequirePackage{wrapfig}
\RequirePackage{verbatim}

\RequirePackage[colorlinks=true,linkcolor=blue,urlcolor=blue,hyperfootnotes=false,citecolor=black]{hyperref}
\RequirePackage{hypernat} % makes hyperref play nice with natbib
\allowdisplaybreaks[1]
\RequirePackage[capitalise]{cleveref}

\newcommand{\Tone}{T_{1}}
\newcommand{\Ttwo}{T_{2}}
\newcommand{\Ttwostar}{T_{2}^{\mathrm{*}}}
\newcommand{\Techo}{T_{2}^{\mathrm{echo}}}
\newcommand{\Gecho}{\Gamma_{\phi}^{\mathrm{echo}}}
\newcommand{\Gstar}{\Gamma_{\phi}^{\mathrm{*}}}
\newcommand{\Gphi}{\Gamma_\phi}
\newcommand{\Gqp}{\Gamma_\mathrm{qp}}
\newcommand{\xqp}{x_\mathrm{qp}}

\newcommand{\EJ}{E_{\mathrm{J}}}
\newcommand{\EC}{E_{\mathrm{C}}}

\newcommand{\EJone}{E_{\mathrm{J1}}}
\newcommand{\EJtwo}{E_{\mathrm{J2}}}
\newcommand{\EJzero}{E_{\mathrm{J0}}}

\newcommand{\sens}{\left \vert\nicefrac{d\fzeroone}{d\Bperp}\right \vert}

\def\bra#1{\mathinner{\langle{#1}|}}
\def\ket#1{\mathinner{|{#1}\rangle}}

\newcommand{\fzeroone}{f_{01}}
\newcommand{\fzerotwoovertwo}{f_{02}/2}

\newcommand{\fcav}{f_{\mathrm{c}}}

\newcommand{\Bperp}{B_{\mathrm{\perp}}}
\newcommand{\Bparone}{B_{\mathrm{\parallel}, 1}}
\newcommand{\Bpartwo}{B_{\mathrm{\parallel}, 2}}
\newcommand{\Bpar}{B_{\mathrm{\parallel}}}

\newcommand{\Bx}{B_{\mathrm{x}}}
\newcommand{\By}{B_{\mathrm{y}}}
\newcommand{\Bz}{B_{\mathrm{z}}}

\newcommand{\Bcrit}{B^{\mathrm{crit}}}
\newcommand{\Bcritperp}{B_{\mathrm{\perp}}^{\mathrm{crit}}}
\newcommand{\Bcritparone}{B_{\mathrm{\parallel}, 1}^{\mathrm{crit}}}
\newcommand{\Bcritpar}{B_{\mathrm{\parallel}}^{\mathrm{crit}}}

\newcommand{\Bphinaught}{B_{\Phi_{0}}}
\newcommand{\Bphinaughtsquid}{B_{\Phi_{0},\mathrm{SQUID}}}

\newcommand{\sinc}{\mathrm{sinc}}
\begin{document}

\title{Magnetic-field resilience of 3D transmons with thin-film Al/AlO$_x$/Al Josephson junctions approaching 1 T}

\author{J.~Krause}
\altaffiliation{These authors contributed equally to this work.}
\affiliation{Physics Institute II, University of Cologne, Zülpicher Str. 77, 50937 Köln, Germany}
\author{C.~Dickel}
\altaffiliation{These authors contributed equally to this work.}
\affiliation{Physics Institute II, University of Cologne, Zülpicher Str. 77, 50937 Köln, Germany}
\author{E.~Vaal}
\affiliation{Physics Institute II, University of Cologne, Zülpicher Str. 77, 50937 Köln, Germany}
\affiliation{JARA Institute for Quantum Information (PGI-11), Forschungszentrum Jülich, 52425 Jülich, Germany}
\author{M.~Vielmetter}
\author{J.~Feng}
\author{R.~Bounds}
\affiliation{Physics Institute II, University of Cologne, Zülpicher Str. 77, 50937 Köln, Germany}
\author{G.~Catelani}
\affiliation{JARA Institute for Quantum Information (PGI-11), Forschungszentrum Jülich, 52425 Jülich, Germany}
\author{J.~M.~Fink}
\affiliation{Institute of Science and Technology Austria, Klosterneuburg 3400, Austria}
\author{Yoichi~Ando}
\email[correspondence should be adressed to: \newline]{dickel@ph2.uni-koeln.de, \newline ando@ph2.uni-koeln.de}
\affiliation{Physics Institute II, University of Cologne, Zülpicher Str. 77, 50937 Köln, Germany}

\begin{abstract}
Magnetic-field-resilient superconducting circuits enable sensing applications and hybrid quantum-computing architectures involving spin or topological qubits and electro-mechanical elements, as well as studying flux noise and quasiparticle loss. 
We investigate the effect of in-plane magnetic fields up to \SI{1}{\tesla} on the spectrum and coherence times of thin-film 3D aluminum transmons. 
Using a copper cavity, unaffected by strong magnetic fields, we can solely probe the magnetic-field effect on the transmons. 
We present data on a single-junction and a SQUID transmon, that were cooled down in the same cavity. 
As expected, transmon frequencies decrease with increasing fields, due to a suppression of the superconducting gap and a geometric Fraunhofer-like contribution. 
Nevertheless, the thin-film transmons show strong magnetic-field resilience:
both transmons display microsecond coherence up to at least \SI{0.65}{\tesla}, and $\Tone$ remains above \SI{1}{\micro\second} over the entire measurable range.
SQUID spectroscopy is feasible up to \SI{1}{\tesla}, the limit of our magnet.
We conclude that thin-film aluminum Josephson junctions are a suitable hardware for superconducting circuits in the high-magnetic-field regime.
\end{abstract}

\maketitle
\def\thefootnote{*}\footnotetext{These authors contributed equally to this work}\def\thefootnote{\arabic{footnote}}

\section{Introduction}

Josephson junctions (JJ) based on aluminum and its oxide (Al/AlO$_x$/Al) have three key properties that have made them the workhorse of circuit QED (cQED)~\cite{Blais04}.
They are routinely fabricated to high quality;
their Josephson energy $\EJ$ can be estimated from the room temperature resistance~\cite{Ambegaokar63};
the $\EJ$ can be controlled with high yield to specifications~\cite{Kreikebaum20} using standard electron beam lithography, and even tuned post fabrication~\cite{Muthusubramanian19,Hertzberg20}.
These properties have enabled various advances in quantum engineering, for example, the scaling up of quantum processors to $\sim$50 qubits~\cite{Arute19} and the fabrication of sophisticated Josephson parametric amplifiers~\cite{Macklin15}.
The cQED framework allows for elucidating the quantum mechanical interaction of various systems with photons, enabling us to understand those systems from a new perspective.
As standard JJ circuits continue to advance, cQED is also applied to more exotic systems like non-conventional JJs, mechanical elements, magnons, quantum dots, spin qubits or Majorana zero modes~\cite{Clerk20}.

When the cQED methods are to be applied to phenomena or systems requiring strong magnetic fields, the magnetic-field compatibility of components used in cQED becomes an issue.
So far, this issue has been explored as need arose.
One component that is particularly useful is a superconducting quantum interference device (SQUID):
Two JJs in parallel form a SQUID.
They are an important tool in, e.g., metrology~\cite{Clarke06}, and a key building block in many cQED quantum computing architectures~\cite{DiCarlo09,Chen14,Reagor18}.
Compatibility of a SQUID with high magnetic fields enables, e.g., studying spin ensembles or even single spins. 
In this context, SQUIDs based on constriction junctions have demonstrated operation up to \SI{6}{\tesla}~\cite{Chen10}.
There is currently a lot of interest in using SQUIDs in external magnetic fields to couple mechanical oscillators~\cite{Rodrigues19,Kounalakis19,Schmidt20,Bera21,Luschmann21}.
Magnetic fields are also a requirement for integrating many spin-qubit architectures with cQED~\cite{Samkharadze18,Mi18}, and for many Majorana zero mode realizations~\cite{Lutchyn10,Cook11} where cQED methods could be used for the readout~\cite{Hassler11}. 
To couple to quantum dots and topological qubits, magnetic field resilient superconducting resonators have been realized~\cite{Samkharadze16,Kroll19,Borisov20}.
But the exploration of superconducting qubits in magnetic fields has so far largely relied on semiconductor nanowire JJs~\cite{Luthi18,Pita-Vidal20,Kringhoj21} or graphene JJs~\cite{Kroll18}.
A notable exception is Ref.~\onlinecite{Schneider19}, which explored a standard single-junction Al/AlO$_x$/Al JJ in a magnetic field, but the findings would suggest that coherence times are already severely limited at \SI{20}{\milli\tesla} of in-plane field.

In this article, we explore the magnetic field dependence of the Josephson energy $\EJ$, and the coherence of transmon qubits~\cite{Koch07} with standard Al/AlO$_x$/Al JJs in a 3D copper cavity~\cite{Paik11,Rigetti12}.
The 3D copper cavity is essentially unaffected by the magnetic field and thus allows for exploring the magnetic field dependence of the transmon without additional complications.
Planar superconducting resonators are themselves vulnerable to magnetic fields, which proved to be a limiting factor in Ref.~\onlinecite{Schneider19}.
We first show the spectrum as a function of out-of-plane magnetic fields and demonstrate how the limitations imposed on qubit coherence by vortices require precise alignment.
With the use of a vector magnet, we can align the magnetic-field axis to the sample plane with high precision.
Thus, we can measure the transmon spectrum and coherence as a function of exact in-plane magnetic field.
We track the transition frequencies of the transmons over a range of $\sim$1--\SI{7}{\giga\hertz} in in-plane magnetic fields of up to \SI{1}{\tesla}.
Based on the spectrum, we try to understand the geometric effects and the magnetic field dependence of the superconducting gap.
One of the transmons has a SQUID loop; therefore, we can investigate combining very sensitive SQUIDs with large magnetic fields.
Overall, even the SQUID transmon maintains sufficient coherence for many of the applications mentioned above.
Thus, we show that Al/AlO$_x$/Al JJs can be operated in high magnetic fields to give coherent qubits matching the demonstrated field compatibility of non-standard SNS JJs~\cite{Kringhoj21}.

\section{Experimental setup}

\begin{figure}
  \begin{center}
  \includegraphics[width=\columnwidth]{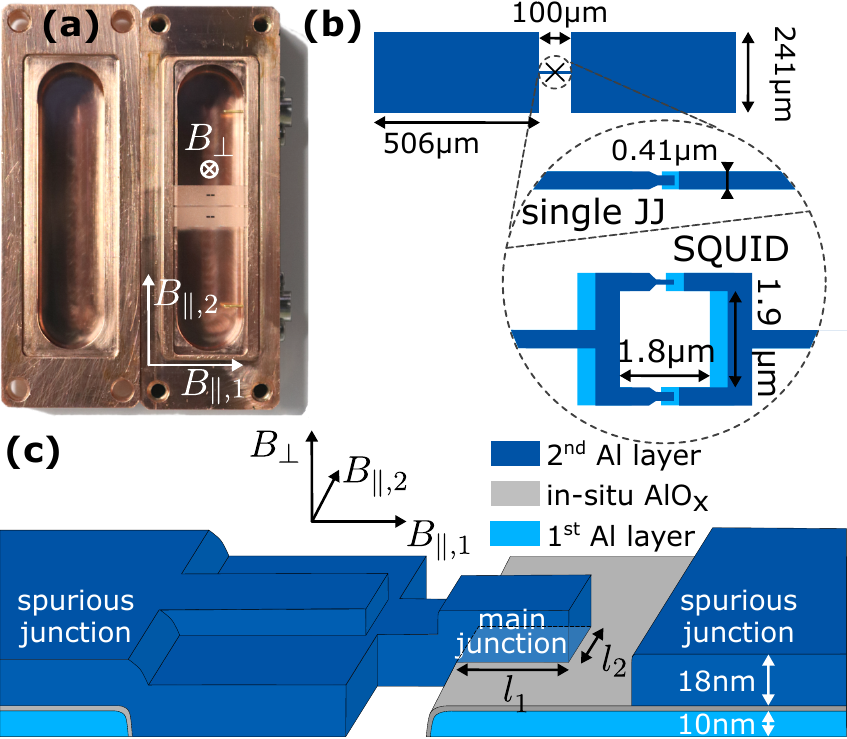}
  \end{center}
  \caption{
    \textbf{(a)} 3D copper cavity with two transmons, referred to as single Josephson junction (JJ) and SQUID.
    \textbf{(b)} Transmon top view with zoom-in on the junction region for both the single JJ and the SQUID device.
    \textbf{(c)} Sketch of a Dolan-bridge JJ relating the magnetic field coordinate system ($\Bparone$,$\Bpartwo$,$\Bperp$) to the JJ geometry.
    }
  \label{fig:fig1}
\end{figure}

In \cref{fig:fig1}, we display the 3D copper cavity containing both transmons, the transmon geometries and a sketch relating the JJ geometry to the magnetic field axes.
The cavity design is based on Ref.~\cite{Abdumalikov13}.
There is one asymmetric SQUID transmon and one single-JJ transmon;
each have their own merits:
On the one hand, the SQUID transmon is sensitive to \si{\micro\tesla} out-of-plane fields $\Bperp$, allowing a precise alignment of the magnetic field parallel to the device plane.
It is also tunable, meaning measurements will cover a wide frequency range allowing to estimate frequency effects at a similar magnetic field.
On the other hand, the single-JJ transmon is less sensitive to flux noise and to the magnetic-field misalignment, and thus it serves as a control device for the SQUID transmon.
As shown in \cref{fig:fig1} \textbf{(b)}, the 3D transmons have long narrow leads to the JJ, making it vortex resilient, even though the big capacitor pads do not have intentional vortex trapping sites. 
Having no magnetic shields, we opted for a small SQUID loop area of \SI{3.4}{\micro\meter\squared}.

The JJs are made with a standard Dolan-bridge design~\cite{Dolan77} with double-shadow evaporation, but for field compatibility we chose a thickness of only \SI{10}{\nano\meter} for the first aluminum layer and \SI{18}{\nano\meter} for the second layer.
The JJs presented were made in the same fabrication run; scanning electron micrographs of the junction region can be found in \cref{sec:device_fab_geometry}.
The design leads to large spurious JJs (see \cref{fig:fig1} \textbf{(c)}) between the two superconducting films, which could complicate the in-plane magnetic field dependence~\cite{Schneider19}.
For more detailed information on the device and on the experimental setup see Ref.~\onlinecite{SOM}.

\section{Out-of-plane magnetic field dependence}
\label{sec:oop_magnetic_field_dependence}

\begin{figure}
  \begin{center}
  \includegraphics[width=\columnwidth]{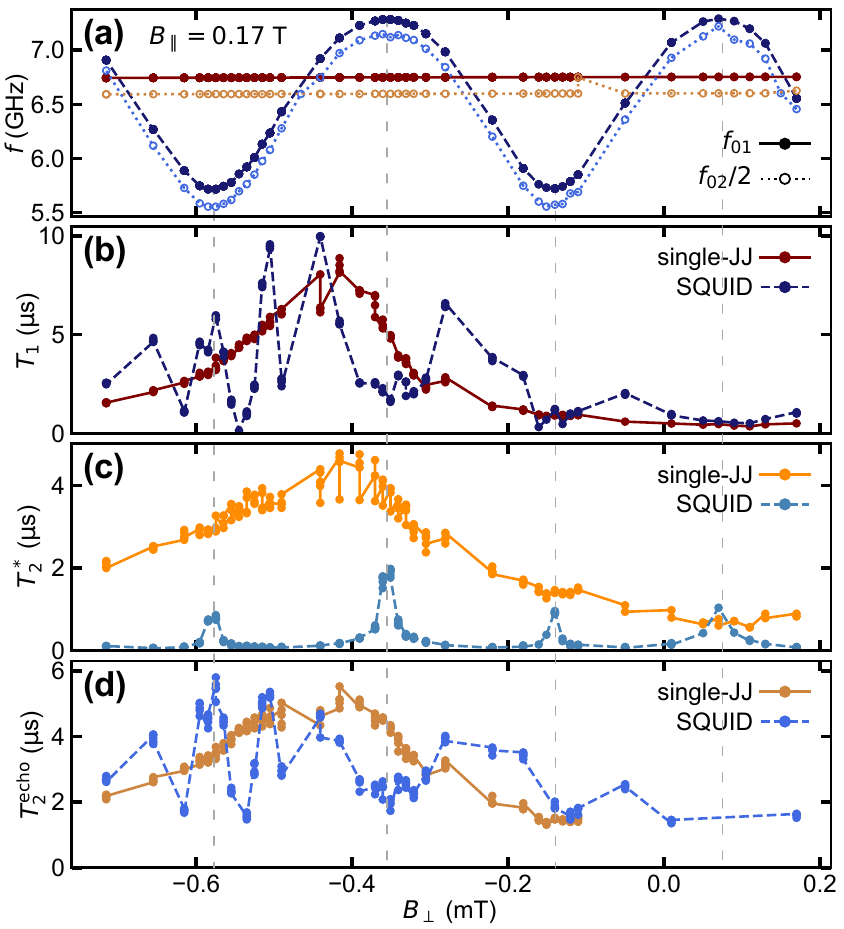}
  \end{center}
  \caption{ Out-of-plane magnetic field dependencies of both transmons, shown here for $\Bparone=\SI{0.17}{\tesla}$.
   \textbf{(a)} First and second transmon transitions $\fzeroone$ and $\fzerotwoovertwo$.
   The SQUID transmon changes with out-of-plane magnetic fields $\Bperp$ which thread the SQUID loop.
   As its constituting JJs are asymmetric, the frequencies oscillate between top and bottom limits, the sweetspots.
   \textbf{(b)} Energy relaxation time $\Tone$.
   High $\Tone$ times are found in a $\Bperp$-interval from \SIrange{-0.7}{-0.25}{\milli\tesla}, deviating from the nominal $\Bperp=0$ based on the SQUID alignment procedure.
   \textbf{(c)} Ramsey dephasing time $\Ttwostar$.
   At the sweetspots, the SQUID frequency is less sensitive to flux noise and $\Ttwostar$ is enhanced.
   \textbf{(d)} Echo dephasing time $\Techo$.
   In the interval of high $\Tone$, $\Techo$ is generally not $2\Tone$-limited for both transmons.
   }
  \label{fig:fig2}
\end{figure}

For every in-plane magnetic field shown in this paper, we sweep the out-of-plane magnetic field, $\Bperp$.
In contrast to Ref.~\onlinecite{Wang14}, where vortex-quasiparticle interplay is explored, we do not perform field cooling; rather, we change the magnetic field with the sample remaining at base temperature, as in Ref.~\onlinecite{Ku16}.
The out-of-plane-field datasets at different in-plane fields are qualitatively similar, even at the highest fields where all quantities can be measured.
As an example, \cref{fig:fig2} \textbf{(a)} shows two-tone spectroscopy peaks of the transmon frequencies for a $\Bperp$ range of $\sim$\SI{1}{\milli\tesla} at $\Bparone=\SI{0.17}{\tesla}$.
We always measure both the first and the second excitation energies of the transmons, $\fzeroone$ and $\fzerotwoovertwo$, to be able to estimate the $\EJ$ and the charging energy $\EC$.~\cite{Luthi18,Schneider19}.
While the frequency of the single-JJ transmon changes only $\sim$\SI{10}{\mega\hertz} over this range in $\Bperp$, the SQUID transmon frequency oscillates between top and bottom limits, the sweetspots.
The sweetspots are determined by the sum and difference of the $\EJ$s of the constituent JJs.
The models for the $\Bperp$ dependence of the spectrum for both transmons can be found in \cref{sec:EC_EJ_estimation}.

We also measure the relaxation time $\Tone$ and the Ramsey and Hahn-echo dephasing times $\Ttwostar$ and $\Techo$ of the transmons for different $\Bperp$ at a fixed $\Bparone$.
One can see in \cref{fig:fig2} \textbf{(b)}-\textbf{(d)} that both transmons show higher coherence at a finite $\Bperp$ (around $\Bperp\simeq \SI{0.4}{\milli\tesla}$).
The $\Bperp=0$ point is based on aligning the parallel magnetic field based on the SQUID oscillation (\cref{sec:Magnet_Alignment}). 
Therefore we have to sweep $\Bperp$ at every in-plane magnetic field and map out the value at which $\Tone$ is maximized; we call this $B_0$ and consider it to be an offset in the perpendicular field dependence.
The offset $B_0$ seems to follow a roughly linear trend as a function of $\Bparone$.
This is an interesting observation, likely related to vortex physics, but we do not have a concrete understanding at this point (for more details, see \cref{sec:Bperp_dependence_T1_fmax}).

Apart from the existence of $B_0$, to understand the effect of $\Bperp$  on $\Tone$, we consider loss due to superconducting vortices coupling to the transmon current (see \cref{sec:vortex_loss} for details) and the Purcell limit imposed by the cavity (see \cref{sec:timedomain_limits}). 
For the single-JJ transmon, the frequency remains practically constant when sweeping $\Bperp$, thus the change in $\Tone$ is likely due to vortices.
The loss scales linearly with $\Bperp-B_0$ sufficiently far away from the maximum $\Tone$, but the onset of vortex loss is not linear.
For the SQUID transmon, we consistently find $\Tone$ to be lower at the top sweetspot than at the bottom sweetspot.
Looking at the frequency dependence of $\Tone$, we find that for high frequencies it is Purcell limited (see \cref{sec:timedomain_limits}).
The $B_0$ values for a given $\Bparone$ are similar for both transmons (see also \cref{sec:Magnet_Alignment}).

The dephasing times for both transmons do not reach $2\Tone$ in the high $\Tone$ interval.
Close to the cavity resonance frequency, photon shot noise from the cavity is a limiting factor to $\Techo$ (see \cref{sec:timedomain_limits}).
Compared to the single-JJ transmon the SQUID shows drastically reduced $\Ttwostar$ with a clear sweetspot enhancement.
For the $\Techo$, the sweetspot enhancement is less clear.
Thus, the SQUID transmon data points to slow noise in $\Bperp$, limiting $\Ttwostar$ but not $\Techo$.

\section{In-plane magnetic field dependence of the spectrum}
\label{sec:ip_magnetic_field_dependence}

\begin{figure}
  \begin{center}
    \includegraphics[width=\columnwidth]{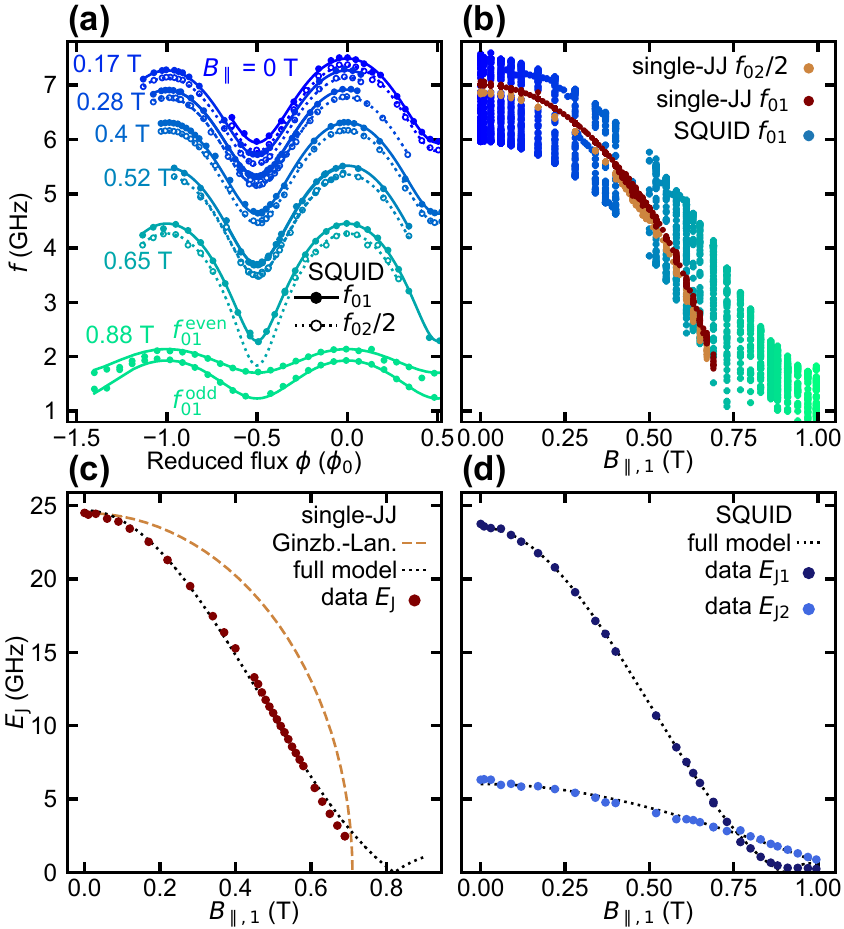}
  \end{center}
  \caption{Spectroscopy for in-plane magnetic fields.
    \textbf{(a)} Examples of the flux dependence of the SQUID transmon frequency:
    For every $\Bparone$, we sweep $\Bperp$ to tune the SQUID transmon.
    We measure $\fzeroone$ and $\fzerotwoovertwo$.
    With increasing $\Bparone$ (color scale corresponds to that in panel \textbf{(b)}), both frequencies decrease, and eventually the $\fzerotwoovertwo$ transition can no longer be measured.
    For high $\Bparone$, $\fzeroone$ is split for even and odd charge parity, which is shown for $\Bparone = \SI{0.88}{\tesla}$.
    \textbf{(b)} SQUID $\fzeroone$ and single-JJ $\fzeroone$, $\fzerotwoovertwo$ transmon transitions versus in-plane magnetic field $\Bparone$.
    \textbf{(c)} extracted Josephson energy $\EJ$ for the single-JJ transmon.
    We correct for different systematic errors, for details see \cref{sec:EC_EJ_estimation}.
    A simple Ginzburg-Landau (GL) theory for the superconducting gap provides neither qualitative nor quantitative agreement.
    Combining GL theory with the flux penetration into an extended junction, \cref{eq:EjInMagneticField}, we obtain better agreement (dotted line).
    \textbf{(d)} Josephson energies $\EJ$ for the two Josephson junctions forming the asymmetric SQUID transmon.
    The larger junction $\EJone$ is consistent with a second Fraunhofer lobe emerging for $\Bparone>\SI{0.9}{\tesla}$.
  }
  \label{fig:fig3}
\end{figure}

Next we consider the in-plane magnetic field dependencies of the two transmons. 
Here we focus on the data obtained for the $\Bparone$ direction.
For every in-plane-field, $\Bparone$, we sweep $\Bperp$ to perform a full set of measurements as explained in \cref{sec:oop_magnetic_field_dependence}.
First, we show how the transmon spectra evolve in parallel magnetic fields.
As one can see in \cref{fig:fig3} \textbf{(b)}, both transmons decrease in frequency at higher magnetic fields.
As the magnetic field increases, the difference between the top and bottom sweetspot frequencies increases (\cref{fig:fig3} \textbf{(a)}), indicating that the $\EJ$s of the two constituting JJs evolve differently.
For high $\Bparone$, we observe large charge-parity splitting due to the decreasing $\nicefrac{\EJ}{\EC}$-ratio.
Thus, for $\Bparone=\SI{0.88}{\tesla}$, the two parity branches of $\fzeroone$ are plotted.

Having measured $\fzeroone$ and $\fzerotwoovertwo$ for both transmons we can estimate $\EJ$ (and $\EC$) as described in \cref{sec:EC_EJ_estimation}.
For the high-field/low-$\nicefrac{\EJ}{\EC}$-ratio regions, charge-parity splitting is used to estimate $\EJ$.
The resulting $\EJ$ as a function of $\Bparone$ is shown in \cref{fig:fig3} \textbf{(c)} and \textbf{(d)} for the single-JJ and SQUID transmon, respectively.
A na\"ive estimate based on the Ginzburg-Landau (GL) theory for the superconducting gap provides neither qualitative nor quantitative agreement for the in-plane field dependence of $\EJ$ (\cref{fig:fig3} \textbf{(c)}).
We therefore combine GL theory with a Fraunhofer term describing the flux penetration into an extended junction
\begin{equation}
   \EJ(\Bpar)=\EJzero \sqrt{1-\left(\frac{\Bpar}{\Bcritpar} \right)^2} \, \left \vert \sinc \left ( \frac{\Bpar}{\Bphinaught} \right )\right \vert,
   \label{eq:EjInMagneticField}
\end{equation}
where $\EJzero$ denotes the Josephson energy at zero field, $\Bcritpar$ the in-plane Ginzburg-Landau critical field, and $\Bphinaught$ the in-plane field for which one superconducting flux quantum ($=\nicefrac{h}{2e}$) threads the JJ.
$\Bphinaught$ is inversely proportional to the in-plane cross section of the junction, defined by its finger width $l_2$ (see \cref{fig:fig1} \textbf{(c)}) and a constant insulator thickness.
As the JJs differ in finger width, each JJ has a different $\Bphinaught$.
Assuming the same critical field $\Bcritpar = \SI{1.03}{\tesla}$ for all three junctions, we find the independently measured junction dimensions are consistent with the estimated $\Bphinaught$ (\cref{tab:FieldPeriodicity}).
Taking the values for $l_2$ and $\Bphinaught$ we can calculate the height of the in-plane cross section threaded by $\Bparone$ which amounts to a plausible \SI{10}{\nano\meter}.
In particular for the asymmetric SQUID transmon this model fits the distinctive behavior of the individual JJs forming the SQUID loop (\cref{fig:fig3} \textbf{(d)}):
The larger JJ shows a rapid decrease in $\EJ$ followed by a slight upturn for $\Bparone>$ \SI{0.9}{\tesla} that is consistent with the emergence of a second Fraunhofer lobe.
The smaller JJ in turn is less affected by flux penetration and its $\EJ$ decreases slowly and monotonically.
While a full BCS modeling of the superconducting gap for thin films could further improve the fits, it is clear that junction geometry plays a role also for conventional Al/AlO$_x$/Al JJs and should be considered when targeting them for operation in high magnetic fields.
The overall JJ footprint should be small and it should especially be narrow in the axis parallel to the magnetic field.

\begin{table}
  \centering
   \begingroup
   \renewcommand{\arraystretch}{1.25} 
   \centering
   \begin{tabular*}{0.8\columnwidth}{c @{\extracolsep{\fill}} ccccc}
      \hline
      & $\EJ(\Bparone=0)$ & $\Bphinaught$ & $l_2$ \\
      \hline
      Single-JJ & \SI{24.7}{\giga\hertz} & \SI{0.83}{\tesla} & \SI{231}{\nano\meter}\\
      SQUID JJ$_1$ & \SI{23.5}{\giga\hertz}  & \SI{0.90}{\tesla}& \SI{206}{\nano\meter}\\
      SQUID JJ$_2$ & \SI{6}{\giga\hertz}  &  \SI{1.65}{\tesla}& \SI{122}{\nano\meter}\\
      \hline
      \vspace{0.01cm}
   \end{tabular*}
   \endgroup
  \caption{
  Parameters of the three JJs. 
  To determine $\Bphinaught$, the in-plane field for which a superconducting flux quantum is threading the JJ, we fit \cref{eq:EjInMagneticField} to the data in \cref{fig:fig3} \textbf{(c)} and \textbf{(d)} assuming the same GL critical field $\Bcritpar=\SI{1.03}{\tesla}$ for all JJs.
  Then $\Bphinaught$ should be inversely proportional to the junction finger width $l_2$, as determined by SEM imaging.}
  \label{tab:FieldPeriodicity}
\end{table}

In \cref{fig:fig3} \textbf{(b)} there is a gap in the SQUID data between \SI{0.4}{\tesla} and \SI{0.5}{\tesla} and the single-JJ data is more noisy in this area.
In this region, no clear SQUID oscillations can be observed when sweeping $\Bperp$.
Measurements of the cavity frequency as a function of $\Bperp$ are not reproducible and the cavity frequency is only stable for several minutes, making qubit spectroscopy for both qubits challenging.
However, the data points that could be gathered for the single-JJ transmon are generally consistent with the data outside this region.
This instability can also be observed when measuring in the $\Bpartwo$ direction, but it arises already at low fields around \SI{20}{\milli\tesla}. 
It is for this reason that we focus on the $\Bparone$ direction here.
Details on these instabilities for $\Bparone$  and $\Bpartwo$ can be found in Ref.~\onlinecite{SOM}.
We suspect spurious JJs inherent in our simple fabrication are responsible;
it would be beneficial to avoid them when exploring large magnetic fields~\cite{Schneider19}.

Eventually our measurements become limited by the decreasing signal-to-noise ratio as the dispersive shifts of the transmons become small.
Therefore we did not measure the single-JJ transmon at magnetic fields above \SI{0.69}{\tesla}.
However, we can measure characteristic SQUID oscillation over the entire field range of \SI{1}{\tesla} that is available to us, as the distinctive frequency modulation helps to identify the SQUID transmon transitions.
Unfortunately, because values of $\Bphinaught$ of the SQUID junctions are above or close to $\Bcritpar$, the upturn in $\EJone$ for $\Bparone>$ \SI{0.9}{\tesla} is relatively weak.

\section{In-plane magnetic field dependence of the coherence times}
\label{sec:qubit_coherence_sensitivity_analysis}

\begin{figure*}
   \begin{center}
      \includegraphics[width=\textwidth]{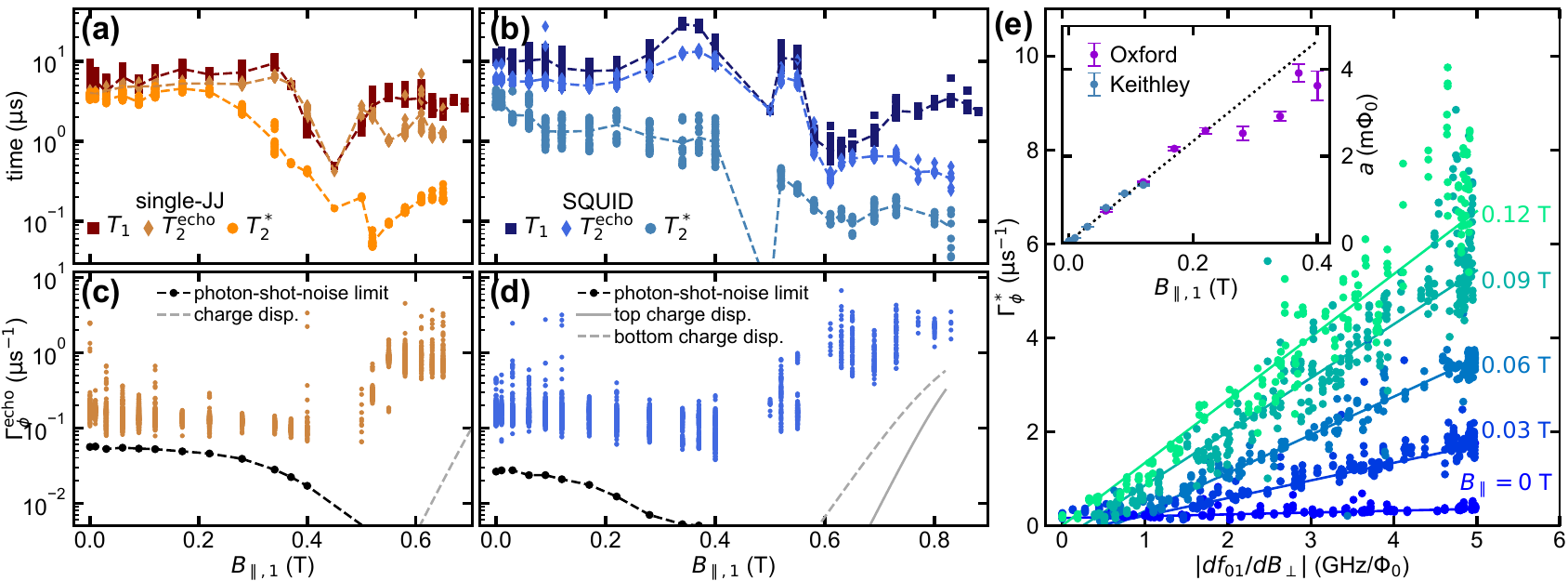}
   \end{center}
   \caption{
      Transmon qubit coherence as a function of $\Bparone$.
      The highest \SI{5}{\percent} of coherence times $\Tone$, $\Techo$, $\Ttwostar$ for \textbf{(a)} single-JJ and \textbf{(b)} SQUID transmon at each $\Bparone$.
      Dashed lines indicate the mean of the high-coherence data at each field.
      Microsecond coherence is maintained up to at least \SI{0.7}{\tesla}, with $\Tone$ above \SI{1}{\micro\second} over the entire measurable range.
      \textbf{(c)} and \textbf{(d)} Pure echo dephasing rates $\Gecho=\nicefrac{1}{\Techo}-\nicefrac{1}{(2\Tone)}$ versus parallel magnetic field $\Bparone$.
      For low magnetic fields (high frequencies) $\Gecho$ is limited by photon shot noise.
      In high magnetic fields the transmons approach the low $\nicefrac{\EJ}{\EC}$ limit and the charge dispersion $\fzeroone(n_g=0)-\fzeroone(n_g=0.5)$ increases, eventually limiting the coherence.
      \textbf{(e)} The pure Ramsey dephasing rate $\Gstar$ as a function of the SQUID frequency sensitivity $\sens$.
      For every in-plane magnetic field $\Bparone$ we observe a linear dependence $\Gstar=a\sens+b$.
      The inset shows the slopes $a$ as a function of $\Bparone$, which suggest that the noise in $\Bperp$ increases linearly with $\Bparone$.
      The observed noise level and trend are independent of the current source connected to the $\Bparone$ magnet coil (named Oxford and Keithley in the legend).
      We believe this noise is caused by mechanical vibrations (see text).
   }
   \label{fig:fig4}
\end{figure*}

Now we turn from the energy spectrum of the transmons to the coherence as a function of $\Bparone$.
At each $\Bparone$, data sets as the one shown in \cref{fig:fig2} were taken.
As coherence times vary, we will first focus on the maximum coherence times measured at each $\Bparone$.
\cref{fig:fig4} (\textbf{a}) and (\textbf{b}) show the highest \SI{5}{\percent} of $\Tone$, $\Ttwostar$ and $\Techo$ for every $\Bparone$.
As seen in \cref{fig:fig2}, the $\Bperp$ for the maximum $\Tone$, $\Ttwostar$ and $\Techo$ do not necessarily coincide.

We observe microsecond $\Tone$ over the entire $\Bparone$-range measurable in time domain.
While the single-JJ transmon $\Tone$ is essentially constant up to \SI{0.4}{\tesla}, the SQUID transmon $\Tone$ shows a slight improvement with a maximum $\Tone$ of more than \SI{30}{\micro\second} for $\Bparone=\SI{0.34}{\tesla}$.
At that point, the perpedicular field offset $B_0$ for maximum $\Tone$ coincides with the bottom sweetspot, and as we noted earlier (see \cref{sec:oop_magnetic_field_dependence}) the bottom sweetspot $\Tone$ is usually longer than at the top sweetspot.
A slight $\Tone$ improvement is also expected because at higher fields and lower frequencies, the Purcell effect is reduced.
In the instability region between \SI{0.4}{\tesla} and \SI{0.5}{\tesla}, the few data points for the single-JJ transmon (and one data point for the SQUID device) suggest a reduction in $\Tone$.
While the single-JJ transmon $\Tone$ stabilizes at a slightly lower 2~-~\SI{4}{\micro\second} after the instability region, the SQUID transmon $\Tone$-dependence for high fields is less clear.
We do not understand the sudden drop in $\Tone$ for the SQUID transmon, nor the gradual improvement in $\Tone$ that follows.
From $\Bparone>0.65$~T onwards we were unable to perform time domain measurements at the bottom sweetspot as the frequency became too low.
Before that our data represents the maximum $\Tone$ across the entire SQUID oscillation; for the highest fields, we lose the lowest frequencies.
Comparing our estimate for the closing of the superconducting gap (\cref{sec:ip_magnetic_field_dependence}) with the qubit lifetimes at high $\Bparone$, we conclude that we are not yet limited by quasiparticle tunneling in the measured range for both qubits~\cite{Catelani11} (see \cref{sec:timedomain_limits} for details).

We now discuss qubit dephasing.
While in general microsecond coherence is maintained up to at least \SI{0.7}{\tesla}, it is clear that $\Techo$ is not $\Tone$-limited.
To better understand the limiting factors, we calculate the pure dephasing rate $\Gphi=\nicefrac{1}{\Ttwo}-\nicefrac{1}{(2\Tone)}$ for both Ramsey and echo experiments.
\cref{fig:fig4} \textbf{(c)} and \textbf{(d)} shows $\Gecho$ as a function of $\Bparone$.
Here we do no longer restrict the discussion to the top \SI{5}{\percent} measured coherence times.
Both devices show a qualitatively and quantitatively consistent trend:
For in-plane magnetic fields up to \SI{0.4}{\tesla}, $\Gecho$ shows a slight decrease, meaning improved coherence.
We partially attribute this effect to photon shot noise in the cavity (see \cref{sec:timedomain_limits}), which limits the transmons less as their frequency decreases with increasing field (dashed line).
The data would suggest an effective cavity temperature of \SI{76}{\milli\kelvin}, which is far above the dilution refrigerator base temperature of $\sim\SI{10}{\milli\kelvin}$.
This could likely be improved by better shielding and filtering.
For fields above \SI{0.52}{\tesla}, we observe increasing qubit dephasing, likely due to charge noise.
The transmons approach the low $\nicefrac{\EJ}{\EC}$ limit and the charge dispersion $\fzeroone(n_g=0)-\fzeroone(n_g=0.5)$ increases (dashed lines).
Here $n_g$ is the charge offset entering the Cooper-pair-box Hamiltonian (see \cref{sec:EC_EJ_estimation}).
With increasing charge dispersion, the transmons become proportionally more sensitive to charge noise~\cite{Koch07}.

As previously noted, the SQUID $\Ttwostar$ shows a strong sweetspot enhancement; we can therefore characterize the noise in $\Bperp$ as a function of $\Bparone$ by performing a sensitivity analysis (see \cref{sec:apdx_sensitivity_analysis}).
Here, pure Ramsey dephasing $\Gstar$ is analyzed as a function of the SQUID frequency sensitivity $\sens$ (\cref{fig:fig4} \textbf{(e)}).
For every in-plane magnetic field $\Bparone$ we observe a linear dependence $\Gstar=a\sens+b$.
The inset shows the slope $a$ as a function of $\Bparone$, which suggest that noise in $\Bperp$ increases linearly with $\Bparone$.
The observed noise level and trend are independent of the current source powering the $\Bparone$ magnet coil; 
we compare the Oxford Instruments Mercury iPS to a low-noise Keithley current source (which cannot reach the currents required for higher fields).
This suggests the noise is not due to the current source of the $\Bparone$ magnet.
A possible explaination could be vibrations in the setup that convert $\Bparone$ to $\Bperp$.
Vibrations are usually low frequency and the noise in $\Bperp$ would increase proportionally with $\Bparone$.
$\Gstar$ would be sensitive to this kind of low-frequency noise.
We attempted to confirm this theory by measuring while turning off the pulse-tube cooler, which is likely the main source of vibrations in the dilution refrigerator, but the turning off led to flux jumps and we could not recalibrate in the time the fridge stayed cold.

A similar analysis was performed for $\Techo$ measurements but $\Gecho$ as a function of $\sens$ is essentially flat, likely because it is mainly limited by photon shot noise or other noise sources that are not $\Bperp$ dependent (see \cref{sec:timedomain_limits}).
Due to the asymmetry of the SQUID, $\sens$ has an upper limit~\cite{Hutchings17}, for a more symmetric SQUID one could increase the $\sens$ until flux noise would become a dominant noise source.
The asymmetry was useful for extracting the magnetic-field dependence of the individual JJs, but for studying flux noise, a symmetric SQUID would be beneficial.
The fact that $\Gecho$ does not show a strong $\Bparone$ dependence is consistent with noise due to mechanical vibrations limiting $\Gstar$, because mechanical vibrations are expected to be low-frequency and the noise can be largely echoed away.
A similar situation is reported in Ref.~\onlinecite{LuthiPhD19}.

\section{Conclusion}

The present results show that for many applications in magnetic fields up to \SI{0.4}{\tesla}, the standard Al-AlO$_x$-Al JJs can be a viable option.
In this regime  $\Tone$ and $\Techo$ times remained largely unaffected in our transmons, but accurate in-plane alignment of the magnetic field is paramount to preserve coherence.
We use thin aluminum films to increase the in-plane critical field and narrow leads to minimize vortex losses.
For higher fields, coherence times are reduced compared to low-field levels, but the standard Al/AlO$_x$/Al transmon can be operated at magnetic fields up to \SI{1}{\tesla}, comparable to semiconductor nanowire transmons~\cite{Kringhoj21}, while exhibiting better coherence times.
For the $\Bparone$ direction, the frequency dependence of the transmon was found to be reasonably well described by a simple model, which considers the gap closing according to the Ginzburg-Landau theory, and a Fraunhofer-like geometrical contribution.
In addition, we have shown that the operation of a SQUID transmon is possible in high in-plane fields, although vibrations of the magnet relative to the sample and noise from magnet current sources could become a limiting factor.
These challenges seem solvable with better vibrational damping of the dilution refrigerator and the use of persistent current magnets. 
However, between \SI{0.4}{\tesla} and \SI{0.5}{\tesla}, regular SQUID oscillations could not be observed and the cavity frequency was unstable.
We speculate this is due to spurious JJs inherent in the Dolan bridge fabrication.

With thinner films and possibly shifting to a JJ fabrication that minimizes spurious JJs, such as Manhattan style JJs~\cite{Potts01} or JJs that are made with two lithography steps~\cite{Wu17}, it would be possible to make an Al-AlO$_x$-Al JJ transmon that can work above \SI{1}{\tesla}.
If the target magnetic field is known in advance and the film properties are largely characterized, one can account for the reduction in $\EJ$ due to suppression of the superconducting gap.
Then, the Al-AlO$_x$-Al JJ advantages of high quality, decent yield and targeting will remain available even in experiments that require high magnetic fields.
In future, it would be interesting to look into charge parity dynamics and thermal excitation in the transmon at higher fields~\cite{Uilhoorn2021}.
Strong in-plane magnetic fields present an additional tuning knob in cQED, which could help understand the physics of the quasiparticles coupling to the transmon.
We also believe that with slight improvements in the setup, it would be possible to measure the effect of magnetic fields on flux noise and shed light on the nature of the spin ensembles that are believed to cause it~\cite{Kumar16}. 

\begin{acknowledgments}
We would like to thank Ida Milow for her internship in the lab and contributions to our code base.
We thank T. Zent and L. Hamdan for technical assistance and D. Fan for help with setting up the aluminum evaporator.
We thank A. Salari, M. Rössler, S. Barzanjeh, M. Zemlicka, F. Hassani, and M. Peruzzo for contributions in the early stages of the experiment.
This project has received funding from the European Research Council (ERC) under the European Union's Horizon 2020 research and innovation program (grant agreement No 741121) and was also funded by the Deutsche Forschungsgemeinschaft (DFG, German Research Foundation) under CRC 1238 - 277146847 (Subproject B01) as well as under Germany's Excellence Strategy - Cluster of Excellence Matter and Light for Quantum Computing (ML4Q) EXC 2004/1 - 390534769

\end{acknowledgments}

\appendix

\section{Device fabrication, geometry and film thickness}
\label{sec:device_fab_geometry}

\begin{figure}
  \begin{center}
    \includegraphics[width=\columnwidth]{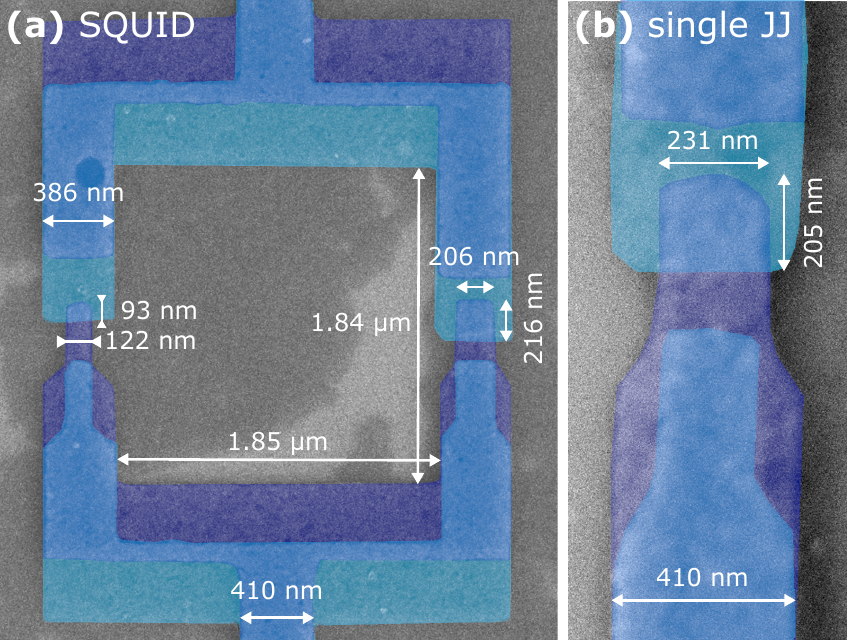}
  \end{center}
  \caption{
    False-colored SEM pictures of \textbf{(a)} the SQUID loop of the SQUID transmon and \textbf{(b)} single JJ of the single-JJ transmon.
    Bottom aluminum layer is overlayed with a turquoise layer, while the top layer is overlayed with a violet layer, leaving the overlap region colored blue.
    Measurements of the different dimensions are indicated (some taken from other images with larger resolution).
  }
  \label{fig:annotated_SEMS}
\end{figure}

The two transmon devices are standard 3D transmons with a Dolan-bridge Josephson junction (JJ)~\cite{Dolan77}.
They were fabricated in a single electron-beam lithography step and a double-shadow evaporation using a Plassys MEB 550S evaporator.
The aluminum has 5N purity. 
To be able to mix and match, many transmon with varying JJ parameters were fabricated in the same run on a large sapphire piece and then diced.
Thus, the two transmons in this experiments, while on two disconnected sapphire pieces, should have very similar aluminum film and junction properties.
The JJ geometry for both with all relevant measurements can be seen in the SEM pictures (\cref{fig:annotated_SEMS}).
Because the taking of SEM pictures alters or destroys the JJs, the actual devices were imaged after measurements were completed.
In the junction test prior to the fabrication of the devices, the relative spread of the room temperature resistances was on the order of \SI{4}{\percent}.
We believe this to  be largely due to the lithography rather than film roughness or a non-uniform oxide layer. 
In the test, 79 out of 96 JJs were working, but we were limited by trying to make small JJs for large SQUID asymmetry.
The reliability of our fabrication process is also confirmed by the fact that the critical current densities of the three junctions studied (proportional to the ratio of $\EJ$s in \cref{tab:FieldPeriodicity} over the junction areas from \cref{fig:annotated_SEMS}) are approximately the same.

Crucially, the film thicknesses for the two evaporations were nominally \SI{10}{\nano\meter} and \SI{18}{\nano\meter} for the bottom and top layers, respectively.
Reducing the film thickness further should be possible using the same evaporator. 
For previous devices with film thicknesses of \SI{15}{\nano\meter} (bottom layer) and \SI{30}{\nano\meter} (top layer), the in-plane critical field was on the order of \SI{250}{\milli\tesla} to \SI{300}{\milli\tesla}.
In contrast, Al films of thickness $d\sim \SI{7}{\nano\meter}$ can remain superconducting up to \SI{3}{\tesla}~\cite{Catelani08}. 
As shown there, at this thicknesses the orbital effect of the parallel field and the Zeeman splitting contribute approximately equally to suppressing superconductivity. 
It is only for thicker films that one can use the relation~\cite{Tinkham04}
\begin{equation}
\Bcritpar = B_c \frac{\sqrt{24} \lambda}{d},
\label{eq:bcritpar_vs_d}
\end{equation}
with $B_c$ the thermodynamic critical field and $\lambda$ the (effective) penetration depth, which qualitatively explain the increase in critical field with decreasing thickness. 
Nonetheless, using the low-temperature value of the critical field for aluminum ($B_c = \SI{10}{\milli\tesla}$), estimating the mean free path $\ell$ to be of the order of the thickness, and using  $\lambda \approx \lambda_L\sqrt{\xi_0/\ell}$, with the London penetration depth  $\lambda_L =\SI{16}{\nano\meter}$ and the coherence length $\xi_0=\SI{1600}{\nano\meter}$, we find from \cref{eq:bcritpar_vs_d} the estimate $\Bcritpar \approx \SI{1}{\tesla}$ for the $d=\SI{10}{\nano\meter}$ thick film in our devices. 
For comparison, the same procedure for $d=\SI{15}{\nano\meter}$ and \SI{30}{\nano\meter} gives $\Bcritpar \approx 0.5$ and 0.2 T, compatible with our measurements. 

We note that the numerical results for the order parameter presented in Ref.~\onlinecite{Catelani08} can be well approximated, not too close to the parallel critical field, by the Ginzburg-Landau formula 
\begin{equation}\label{eq:GL_Gap}
  \Delta(\Bpar)=\Delta_0\sqrt{1-\left(\frac{\Bpar}{\Bcritpar}\right)^2},
\end{equation}
although with a (fitted) critical field larger than the one obtained numerically. 
While this justifies the phenomenological use of \cref{eq:GL_Gap} in analyzing the data, in our devices a further complication arises due to proximity effect between two films of different thickness; however, modeling of this effect is beyond the scope of the present work.

\section{Alignment of magnetic axes to sample}
\label{sec:Magnet_Alignment}

Here we illustrate the alignment procedure to align our magnet axes precisely to the in-plane direction of our sample.
We used the SQUID oscillation offset as a signal, to construct the two in-plane axis $\Bparone$ and $\Bpartwo$ from the physical magnet axes $\Bx$, $\By$, $\Bz$.
In our case the magnet $\Bx$ corresponds roughly to $\Bperp$.
The current source connected to the $\Bx$ coil has a finer resolution and lower noise than the one connected to $\By$ and $\Bz$.
We therefore only used the $\Bx$ coil to correct the extra out-of-plane field caused by $\By$ and $\Bz$ and not vice versa.
This is a simple rotation that we apply in software before setting the values.

To determine the alignment we took a 2D map of the cavity frequency as a function of $\Bx$ and $\By$ (or $\Bz$).
These measurements are fast and we can scan the $\Bx$ field for several $\By$ with few visible jumps.
A linear change in the offset of the SQUID oscillation along the $\Bx$ axis with changing $\By$ is due to an additional out-of-plane component of $\By$.
Then a linear fit is performed to find the misalignment which is then corrected by an additional $\Bx$ field as a function of $\By$.
The resulting axis is our $\Bparone$.
An aligned data set can be seen in \cref{fig:magnet_alignment} \textbf{(a)}, a color plot of the cavity resonance frequency normalized line by line vs $\Bperp$ and $\Bparone$.
The stable offset of the oscillation over a large range of $\Bparone$ suggests that we have aligned our magnetic field axis to better than \SI{0.05}{\degree}.
We determined the initial misalignment to be  \SI{-0.61}{\degree} between the $\By$ and $\Bparone$ axis.
For very low field there is usually a small deviation which we attribute to small residual ferromagnetism in the vicinity of our sample being magnetized.
A more concrete example with misaligned and aligned data for the $\Bpartwo$ direction can be found in Ref.~\onlinecite{SOM}.

\section{Unusual $\Bperp$-dependence of $\Tone$ and the maximum qubit frequency}
\label{sec:Bperp_dependence_T1_fmax}

The alignment of the magnet axes on the SQUID oscillation seems natural and gives a straightforward linear alignment procedure.
While one would expect $\Tone$ as well as the qubit frequency (meaning the superconducting gap) to be maximal at the nominal $\Bperp=0$ (which depends on the alignment), we observed that they take their maximum values at finite values of $\Bperp$; 
furthermore, these values are different for the maximum $\Tone$ and the maximum qubit frequency.

\begin{figure}
  \begin{center}
  \includegraphics[width=\columnwidth]{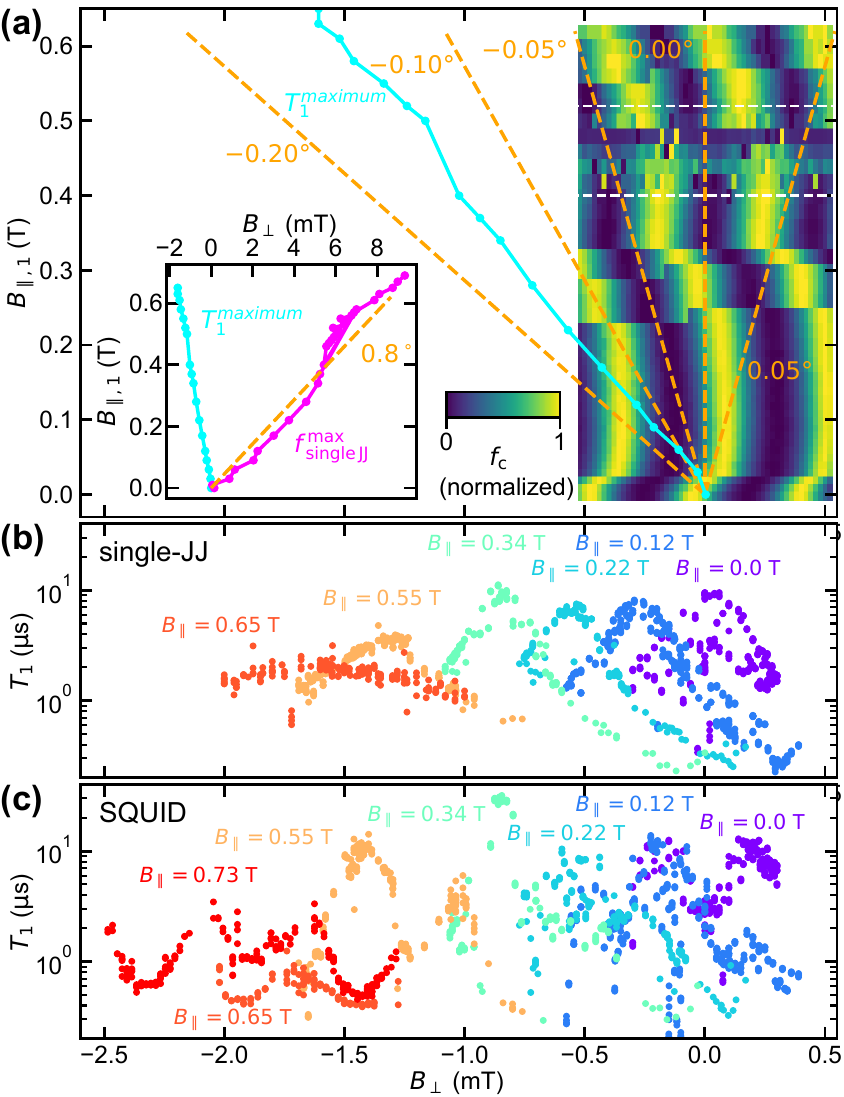}
  \end{center}
  \caption{
    \textbf{(a)} Cavity frequency (line-by-line normalized for contrast) as a function of $\Bperp$ and $\Bparone$ (color plot).
    We observe clear SQUID oscillations in the cavity frequency with a stable period for a large range of magnetic field.
    Occasional jumps can change the flux offset of the oscillations.
    The region between the two dashed white lines shows no stable SQUID oscillations.
    For perfect alignment, the oscillation offset should be constant for different $\Bparone$.
    Orange dashed lines corresponding to different angular misalignment are given as a guide to the eye.
    We conclude that our alignment should be within $\pm$\SI{0.05}{\degree} with respect to the plane of the SQUID.
    The cyan line indicates the ($\Bperp$, $\Bparone$) values corresponding to the maximum $\Tone$ of the single-JJ transmon.
    The maximum $\Tone$ values follow an axis at a $\sim$\SI{-0.15}{\degree} angle with respect to the sample plane.
    The inset additionally shows the ($\Bperp$, $\Bparone$) values corresponding to the estimated maximum frequency of the single-JJ transmon (magenta) which follows an axis that is at a $\sim$\SI{0.8}{\degree} angle with respect to the sample plane.
    \textbf{(b)} and \textbf{(c)} Example data sets of $\Tone$ as a function of $\Bperp$ for single-JJ and SQUID transmon respectively for different $\Bparone$.
  }
  \label{fig:magnet_alignment}
\end{figure}

When looking at the $\Bperp$ corresponding to the largest $\Tone$ at a given $\Bparone$ for both transmons (\cref{fig:magnet_alignment} \textbf{(b)} and \textbf{(c)}, we see that it increasingly deviates from $\Bperp=0$. 
In the following, we focus on the single-JJ transmon $\Tone$ data, because it shows more clear peaks as there is no additional frequency dependence that complicates the picture. 
We designate the $\Bperp$ corresponding to optimal $\Tone$ as $B_0$.
$B_0$ changes linearly with $\Bparone$, such that we can estimate the angle with respect to the sample plane which is roughly \SI{-0.15}{\degree} (data labeled $\Tone^\mathrm{maximum}$ in \cref{fig:magnet_alignment} \textbf{(a)}).
The dependence of $\Tone$ on $\Bperp$ is likely due to vortex creation, which takes place largely in the large capacitor pads.
We show in \cref{sec:vortex_loss}, that apart from this offset it appears that the data is consistent with the vortex hypothesis.
Initially we believed that there could be hysteresis in the vortex system, which could lead to an offset in $B_0$.
So when changing $\Bparone$, we tried to scan $\Bperp$ back and forth approaching the estimated $\Bperp=$ \SI{0}{\milli\tesla} point, a procedure laid out in Ref. \onlinecite{Bothner12}. 
However, this procedure did not make a big difference.
Some data on the hysteresis in $\Bperp$ at $\Bparone=0$ can be found in Ref~\onlinecite{SOM} and while we see hysteresis in the SQUID offset and in the $\Tone$ data, it is not necessarily identical.
Ultimately we found that $B_0$ seemed to be stable for up and downscan in $\Bparone$, therefore it appears that hysteresis does not fully explain the effect.
This effect could be investigated in more detail, we did, e.g., not explore the negative direction in $\Bparone$, but it is beyond the scope of this work.

Peculiarly, the $\Bperp$ value corresponding to the maximum frequency of the single-JJ transmon seems to also linearly deviate from $\Bperp=0$ at different $\Bparone$, corresponding to an angle of \SI{0.8}{\degree} with respect to the in-plane direction (see inset of  \cref{fig:single_JJ_vs_Bperp} \textbf{(a)}).
Two example data sets for the frequency of the single-JJ transmon as a function of magnetic field can be found in \cref{fig:single_JJ_vs_Bperp}.
If one assumes that only $\EJ$ is field dependent, the maximum frequency corresponds to the maximum superconducting gap immediately at the JJ.
Possibly due to flux focusing in the vicinity of the JJ, which has a step in the $\Bparone$ direction, there is an additional angle with respect to the sample plane.
While the differences in angle between the SQUID, the vortex system and the maximum frequency of the single-JJ transmon are small in absolute terms, they are clearly distinguishable in our data.

\section{Vortex loss in out-of-plane magnetic fields}
\label{sec:vortex_loss}

\begin{figure}
   \begin{center}
      \includegraphics[width=\columnwidth]{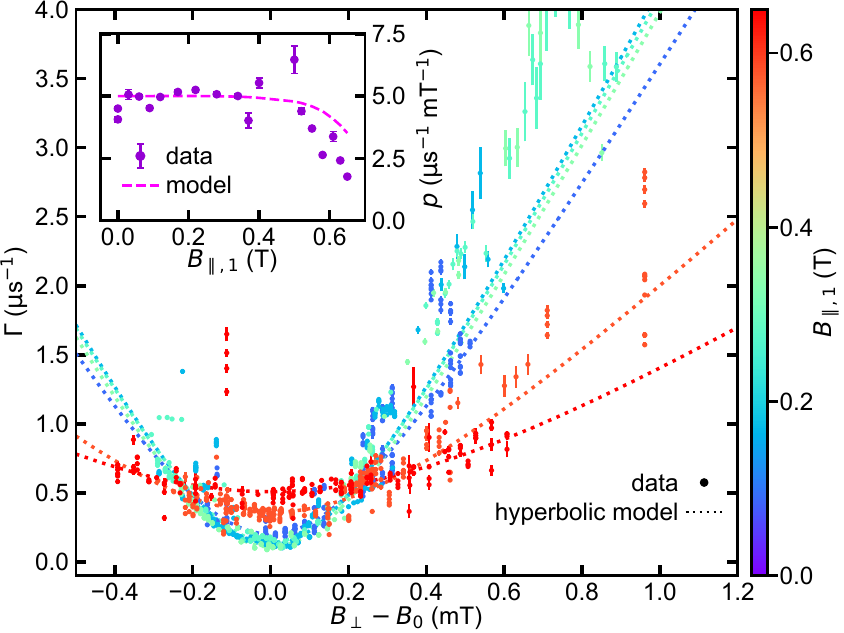}
   \end{center}
   \caption{
      Single-JJ transmon relaxation rate $\Gamma=\nicefrac{1}{\Tone}$ vs $\Bperp-B_0$ for various $\Bparone$.
      The data are fitted using a phenomenological hyperbolic model (\cref{eq:Gammav}.
      The inset shows the fitted asymptotic slope $p$ of \cref{eq:Gammav}.
      Interestingly, for $\Bparone\ge\SI{0.5}{\tesla}$ the slope shows a significant drop for high $\Bparone$.
      We can model this decrease in $p$ (dashed pink line) which we mainly attribute to the decreasing qubit frequency.
   }
   \label{fig:t1_vs_Bperp}
\end{figure}

In \cref{fig:fig4}, we plot the best $T_1$ times as a function of $B_{\parallel,1}$; however, reaching the longest possible $T_1$ crucially depends on finding the appropriate $B_\perp$ value for a given $B_{\parallel,1}$, as we discuss in \cref{sec:Magnet_Alignment}, pointing to the possible role of vortices. 
Indeed, as shown both in resonators~\cite{Song09} and transmons~\cite{Ku16}, the loss is proportional to the number of vortices; above a certain threshold field $B_\mathrm{th}$, this number increases linearly with $B_\perp$.
However, vortices can enter into the large transmon capacitor pads already at fields smaller than $B_\mathrm{th}$~\cite{Song09,Stan04}, leading to a more gradual onset of vortex dissipation. 
To phenomenologically capture this behavior, we fit the vortex contribution to dissipation $\Gamma_\mathrm{v}$ with the formula
\begin{equation}\label{eq:Gammav}
    \Gamma_\mathrm{v} = \sqrt{p^2 \tilde{B}_\perp^2+q^2}-q\, ,
\end{equation}
where $p$ and $q$ are fit parameters, which we discuss below, and $\tilde{B}_\perp = B_\perp - B_0$, with the offset $B_0$ being the value of the perpendicular field where $T_1$ is the largest for a given $B_{\parallel,1}$ (see \cref{sec:Bperp_dependence_T1_fmax}).
We show in \cref{fig:t1_vs_Bperp} the total relaxation rate $\Gamma=1/T_1$ as function of $B_\perp - B_0$ for several values of the parallel field; note that $\Gamma = \Gamma_0 + \Gamma_\mathrm{v}$ includes also the non-vortex contribution $\Gamma_0$. 
Data over a wider range of perpendicular field, showing more clearly a regime of linear dependence of $\Gamma_\mathrm{v}$ on $B_\perp$, is reported in Ref. \onlinecite{SOM}.

In fitting the data of \cref{fig:t1_vs_Bperp}, we fix $q=\SI{1.3}{\per\micro\second}$, while we treat $p$ as a parallel field-dependent quantity. 
The inset in \cref{fig:t1_vs_Bperp} presents the value of $p$ as a function of $B_{\parallel,1}$; 
The coefficient $p$ is the slope in the linear regime of $\Gamma_\mathrm{v}$ vs $B_\perp$. 
As discussed in Ref.~\cite{Song09}, the value of the slope is affected by the so-called flux-flow viscosity $\eta$ and the presence of pinning centers that can lead to vortex creep. 
Considering the model for the flow resistivity of Ref.~\cite{Song09} (see also~\cite{Pompeo08}), we can write
\begin{equation}\label{eq:slope}
    p = p_0 \frac{1}{\sqrt{1-(B_\parallel/B^\mathrm{crit}_\parallel)^2}}\frac{F(f(B_\parallel)/f_d,\epsilon)}{F(f(0)/f_d,\epsilon)} \, ,
\end{equation}
where by construction $p_0$ is the slope at zero parallel field, $f(B_\parallel)$ is the transmon frequency as function of the parallel field, $f_d$ is the depinning frequency, marking the crossover from elastic to viscous response of the vortices, and $0 \le \epsilon \le 1$ is the dimensionless creep parameter. 
For aluminum, the latter two quantities take the values $f_d=\SI{4}{\giga\hertz}$ and $\epsilon=0.15$~\cite{Song09}. 
The function $F$ is defined as
\begin{equation}\label{eq:function_F}
    F(x,\epsilon) = \frac{\epsilon + x^2}{1+x^2} \, .
\end{equation}
Finally, the factor in the middle of \cref{eq:slope} arises as follows: the loss is inversely proportional to the viscosity $\eta$, and the latter is proportional to the upper critical field  $B_{c2} = \Phi_0/(2\pi\xi^2)$, where $\xi \approx \sqrt{\hbar D/\Delta}$ is the coherence length in a disordered superconductor, with $D$ the diffusion constant (physically, the loss increases with the square of coherence length because the latter determines the radius of the vortex core). 
Therefore, we expect $p\propto 1/\Delta(B_\parallel)$, a factor that we estimate using \cref{eq:GL_Gap}.

The curve in the inset of \cref{fig:t1_vs_Bperp} has been plotted using \cref{eq:slope}, with the qubit frequency obtained from the data in \cref{fig:fig3} \textbf{(b)} and $\Bcritparone=\SI{1.03}{\tesla}$, see the caption of \cref{tab:FieldPeriodicity}. 
Hence $p_0=\SI{5}{\per\micro\second\per\milli\tesla}$ is the only free parameter, which has been fixed by fitting the data for $B_\parallel \le \SI{0.4}{\tesla}$; for comparison, accounting for their different frequencies through the function F of \cref{eq:function_F}, the two qubits measured in Ref. \onlinecite{Ku16} have $p_0=0.5$ and \SI{1.2}{\per\micro\second \per \milli\tesla}. 
The curve captures the experimental drop of the slope with parallel field, implying that the decrease in dissipation at low frequency due to pinning has a stronger effect compared to the increase due to the expansion of the vortex cores. 
Based on this result, we expect that by introducing pinning sites or vortex-trapping holes in the pads, the qubit can be made more robust to out-of-plane fields and less sensitive to misalignment, although care must be taken in not increasing dielectric losses~\cite{Chiaro16}.

Returning now to \cref{eq:Gammav}, the parameter $q$ can be related to the threshold field by $B_\mathrm{th} \sim q/p$; however, this identification is meaningful only at zero parallel field, since at higher field (and hence lower frequency) $p$ is suppressed due to pinning. 
In this way, we estimate $B_\mathrm{th} \sim q/p_0 \approx \SI{0.26}{\milli\tesla}$, similar to the value at which decrease in $\Tone$ starts in Ref.~\onlinecite{Ku16}. 
In that case, this value is related there to the lower critical field for vortex entry into a region of the capacitor pads, close to the JJs, of lateral size $\sim \SI{10}{\micro\meter}$. 
However, this explanation is not applicable to our device, since there are no features with comparable dimensions, and we expect vortex entry in the pads already at a few $\SI{}{\micro\tesla}$.
We speculate that $B_\mathrm{th}$ could be related to the number of vortices exceeding the number of pinning sites.
We do not expect vortices to enter the thin leads to the JJs in our device, because the lead width $w=\SI{410}{\nano\meter}$ is only a few times the coherence length $\xi\sim 0.85\sqrt{\xi_0 \ell} \approx\SI{108}{\nano\meter}$ (see \cref{sec:device_fab_geometry}). 
In fact, an order-of-magnitude estimate for the field of vortex entry into the leads $B_\mathrm{v}$ applicable in the case $w \gg \xi$ is $B_\mathrm{v} = \Phi_0/w^2 \approx \SI{10}{\milli\tesla}$~\cite{Stan04}. 
Although the condition $w\gg \xi$ is not satisfied, this value suggests that vortices are not present in the leads in the few mT range of perpendicular field explored in this work.

\section{Estimation of $\EJ$ and $\EC$ from spectroscopy data}
\label{sec:EC_EJ_estimation}

In \cref{fig:fig3} we show $\EJ$ as a function of $\Bparone$.
Here we want to elaborate on how we estimate $\EJ$ and $\EC$ from the measured transmon spectrum.
We also consider systematic errors, such as additional dependence on $\Bperp$ and cavity dressing.

\begin{figure}
   \begin{center}
      \includegraphics[width=\columnwidth]{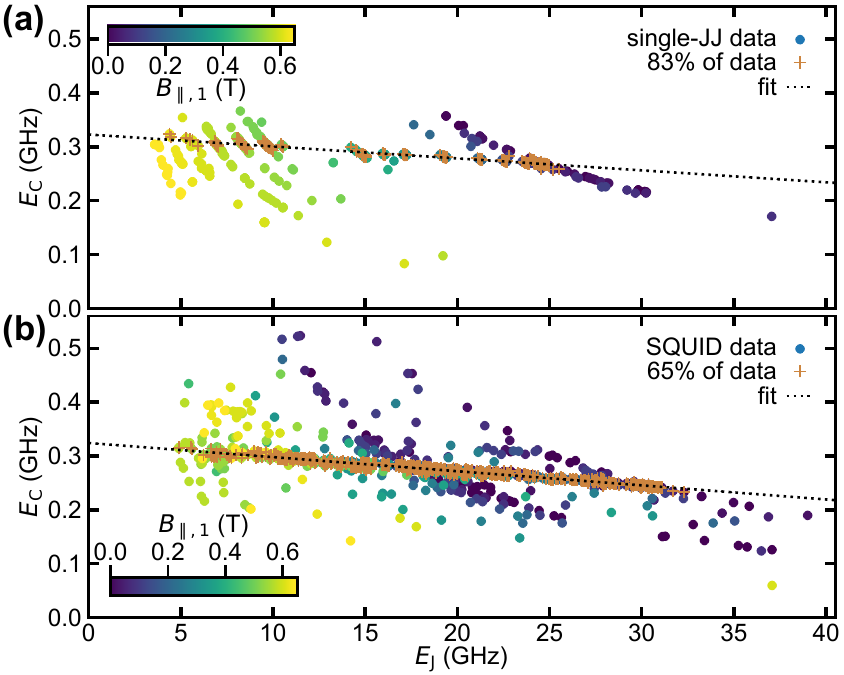}
   \end{center}
   \caption{
      \textbf{(a)} Single-JJ and \textbf{(b)} SQUID transmon $\EC$ as a function of $\EJ$.
      This data was extracted from pairwise spectroscopic measurements of $\fzeroone$ and $\fzerotwoovertwo$.
      For every pair, we fit the transitions to a numerical Cooper-pair box Hamiltonian, giving a value for $\EJ$ and $\EC$ (see text).
      We find a clear correlation of $\EC$ and $\EJ$: \SI{65}{\percent} of all SQUID data and \SI{80}{\percent} of all single JJ qubit data gather around a linear trend to within \SI{10}{\mega\hertz}.
      The outliers can be due to a number of effects: bad peak fits (e.g. picking a wrong photon number peak), flux or $n_g$ jumps between the $\fzeroone$ and $\fzerotwoovertwo$ measurements.
      Fitting a linear dependence we can infer $\EC$ from $\EJ$.
   }
   \label{fig:EC_vs_EJ}
\end{figure}

$\EJ$ and $\EC$ can be extracted from $\fzeroone$ and $\fzerotwoovertwo$ by fitting the measured transitions to a numerical Cooper-pair box Hamiltonian in the charge basis
\begin{equation} \label{eq:transmon_hamiltonian}
\begin{split}
        H = \; \; \;   4  \EC  & \sum_{n=-k}^k (n-n_g)^2 \ket{n} \bra{n}  \\
               + \frac{1}{2} \EJ & \sum_{n=-k}^k   \left( \ket{n}\bra{{n+1}} + \ket{n+1}\bra{n}\right),  
  \end{split}
\end{equation}
with charge states $\ket{n}$, where $n$ stands for the difference in Cooper pairs between the two islands.
A voltage gate or environmental noise can introduce a charge offset $n_g$.
$k$ is the truncation in the charge basis; we usually truncate at $k=20$, thus include 41 states.
That way we obtain accurate results in both the transmon regime and the low $\nicefrac{\EJ}{\EC}$ regime, where $\fzeroone \approx \sqrt{8\EJ \EC}-\EC$ stops being a good approximation. 
Every pair of $\fzeroone$ and $\fzerotwoovertwo$ measurements will then give a value for $\EJ$ and $\EC$.
The data  are shown in \cref{fig:EC_vs_EJ}.

The data shows a clear correlation between $\EJ$ and $\EC$ because the participation of the cavity capacitance depends on the impedance matching between the cavity and transmon mode and therefore on $\EJ$.
The coupling between cavity and transmons is also not constant but depends on $\EJ$ and $\EC$.
The dependence looks very similar for both transmons and we can assume a linear dependence of $\EC$ on $\EJ$.
The outliers in the data set can be due to a number of effects: bad peak fits (e.g. picking a wrong photon number peak), flux or $n_g$ jumps between the $\fzeroone$ and $\fzerotwoovertwo$ measurements or hysteresis in the magnetic field.
Throughout the experiment we used continuous wave spectroscopy.
The powers were constantly adapted as the qubit-cavity detuning grew trying to maintain a balance between visibility and minimizing the shifts due to the readout tone and the AC stark shift.

Both the single-JJ and the SQUID transmon transitions vary with $\Bperp$.
Furthermore, our large spectroscopy dataset has outliers. 
As we sweep a small range in $\Bperp$ for every $\Bparone$, we can identify and reject outliers easily.
To obtain a robust estimate of $\EJ(\Bparone)$, we do not extract all individual values for $\EJ$ from all $\fzeroone$ and $\fzerotwoovertwo$ pairs, but rather fit a model to all transitions measured at a given $\Bparone$ as a function of $\Bperp$.
In the following we will give the models we used for the $\Bperp$ dependence for the two transmons.

For the single-JJ transmon, the out-of-plane field dependence is dominated by a suppression of the superconducting gap. 
As in the case of the in-plane magnetic field, we model it using Ginzburg-Landau dependence of the gap on field~\cite{Tinkham04}
\begin{equation}
   \EJ(\Bperp)\propto\Delta(\Bperp)=\Delta(0)\sqrt{1-\left(\frac{\Bperp}{\Bcritperp}\right)^2}.
   \label{eq:GapVsB}
\end{equation}
Using the same form as in \cref{eq:GL_Gap} seems appropriate since, as discussed at the end of \cref{sec:vortex_loss}, we do not expect vortices to play a role in the leads to the JJ at least up to $B_\perp \sim 10$ mT, which covers the range of perpendicular field in our measurements.
Note that the critical field $\Bcritperp$ of the junction leads should not be confused with the upper critical field $B_{c2}$ of the much wider pads introduced in \cref{sec:vortex_loss}.

\begin{figure}
   \begin{center}
      \includegraphics[width=\columnwidth]{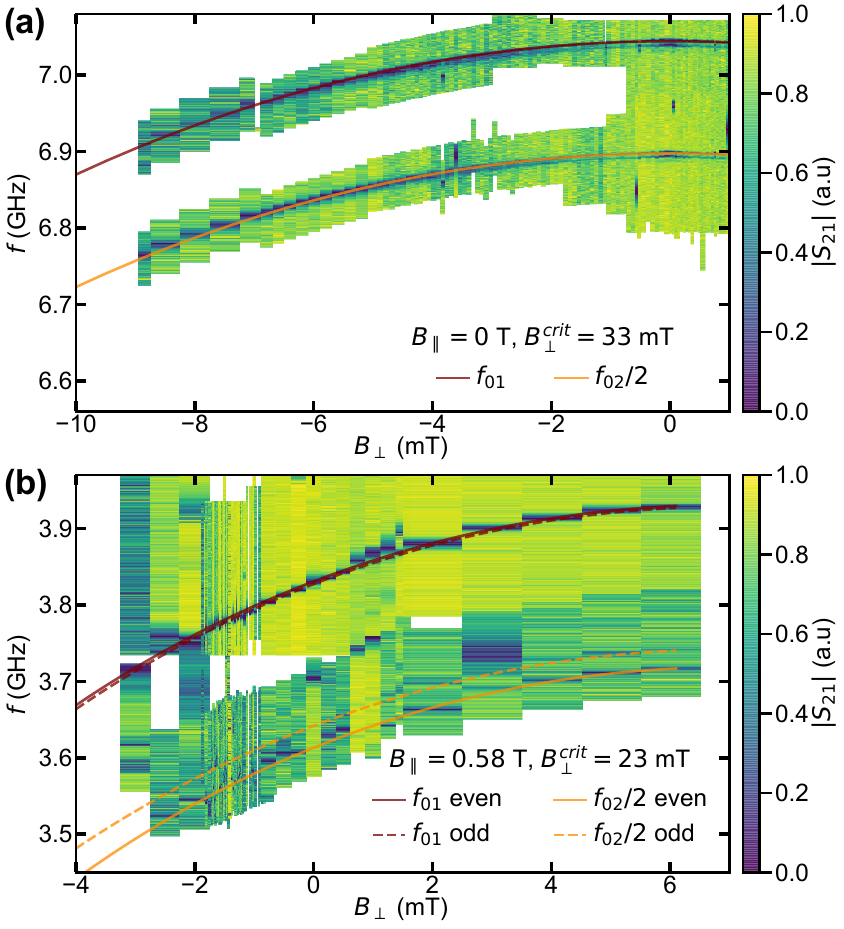}
   \end{center}
   \caption{
      Single-JJ transmon $\fzeroone$ and $\fzerotwoovertwo$ as a function of $\Bperp$ for $\Bparone=\SI{0}{\tesla}$ \textbf{(a)} and for $\Bparone=\SI{0.58}{\tesla}$ \textbf{(b)}.
      The data for \textbf{(a)} was taken in a previous cooldown in a different but nominally identical dilution refrigerator. 
      We use this data to estimate $\Bcritperp$ using a fit of a simple Ginzburg-Landau model.
      The effective $\Bcritperp$ is lower in \textbf{(b)} as the superconductivity is also suppressed by the in-plane field.
   }
   \label{fig:single_JJ_vs_Bperp}
\end{figure}

When simultaneously applying an in-plane and an out-of-plane magnetic field the effective $\Bcritperp(\Bparone)$ is reduced. 
In Ginzburg-Landau theory for thin films, for any angle $\theta$ to the film plane, the critical field $\Bcrit(\theta)$ lies in between $\Bcritperp$ and $\Bcritparone$ and satisfies~\cite{Tinkham04}
\begin{equation}
    \left \vert \frac{\Bcrit(\theta) \sin\,\theta}{\Bcritperp} \right \vert
   +\left ( \frac{\Bcrit(\theta) \cos\,\theta}{\Bcritparone} \right )^2=1.
   \label{eq:BcVsTheta}
\end{equation}
Example data for the single-JJ transmon transitions as a function of $\Bperp$ can be found in \cref{fig:single_JJ_vs_Bperp}.
We only measured spectroscopy for a $\sim\SI{10}{\milli\tesla}$ range in $\Bperp$ at $\Bparone=0$~T and at $\Bparone=0.58$~T.
At other fields, we generally measured a range of $\sim \SI{2}{\milli\tesla}$ in $\Bperp$ around the high-coherence interval, because we want to mainly make the case that high coherence can be maintained.
But the frequency maximum as a function of $\Bperp$ for the single-JJ transmon increasingly deviates from the maximum coherence time  (see \cref{sec:Magnet_Alignment}).
In \cref{fig:single_JJ_vs_Bperp} \textbf{(b)}, the transmon frequency in the high-coherence interval around \SI{-1.5}{\milli\tesla} is about \SI{150}{\mega\hertz} lower than the maximum frequency we measured.
To account for this, we try to estimate the maximum $\EJ$ at every $\Bparone$ by fitting all data we have in $\Bperp$ for this $\Bparone$.
In the fit, we fix $\Bcritperp$ and $\Bcritparone$ and use \cref{eq:BcVsTheta} to extract the effective $\Bcritperp(\Bparone)$ at each $\Bparone$.
The free parameters are the maximum $\EJ$, the $\EC$  and the offset in $\Bperp$.
For the highest fields this suggests a $\sim$\SI{+10}{\percent} correction on the $\EJ$.
Note that when we model the magnetic field dependence of the transmon frequencies, we consider the field dependence of $\EJ$ and therefore of the superconducting gap right at the JJ.
In fact, \cref{eq:GapVsB} accounts for the gap suppression due to the perpendicular component of the field.
This mechanism would result in the first term of \cref{eq:BcVsTheta} to be a square, as the second term, rather than an absolute value; the absolute value originates from the effect of vortices~\cite{Tinkham04}.
As we discussed, it is unclear what ultimately determines the perpendicular critical field in the leads, and hence which formulation is the correct one.
In the measured range of parallel field, the two approaches give effective $\Bcritperp$ differing by at most \SI{25}{\percent}, so we opted to use the well-known \cref{eq:BcVsTheta}.
In future experiments this could be easily explored by measuring larger ranges in $\Bperp$ for each $\Bparone$. 

For the SQUID transmon, the effective $\EJ$ will depend on $\Bperp$ and on the two individual Josephson energies $\EJone$ and $\EJtwo$ according to:
\begin{equation} \label{eq:SQUID_EJ_vs_phi}
\EJ \left ( \Bperp \right ) =  E_{\mathrm{J}, \Sigma} \sqrt{ \alpha_\mathrm{JJ}^2 + \left ( 1 - \alpha_\mathrm{JJ}^2\right ) \cos \left (  \frac{\pi \Bperp}{\Bphinaughtsquid} \right )^2},
\end{equation}
with $E_{\mathrm{J}, \Sigma} = \EJone + \EJtwo $, the JJ asymmetry parameter $ \alpha_\mathrm{JJ} = \nicefrac{\left \vert \EJone - \EJtwo \right \vert}{\left ( \EJone + \EJtwo \right )}$ and the out-of-plane field $\Bphinaughtsquid$ that corresponds to a flux quantum through the SQUID loop.
Intuitively the Josephson energies at the top (bottom) sweetspot correspond to the sum (difference) of the individual $\EJ$s.
Neglecting the suppression of the superconducting gap with $\Bperp$ that we have noted for the single-JJ transmon above, one can fit this dependence to a SQUID oscillation and get a result for $\EJone$ and $\EJtwo$.
In the fitting model, we use the linear relation of $\EJ$ and $\EC$ extracted before.
The suppression of the superconducting gap with $\Bperp$ can be neglected because of the large difference between $\Bcritperp\approx\SI{30}{\milli\tesla}$ and $\Bphinaughtsquid\approx\SI{0.43}{\milli\tesla}$.
We also do not observe that the sweetspot frequencies vary as strongly with $\Bperp$ as the frequency of the single-JJ transmon.

\begin{figure}
   \begin{center}
      \includegraphics[width=\columnwidth]{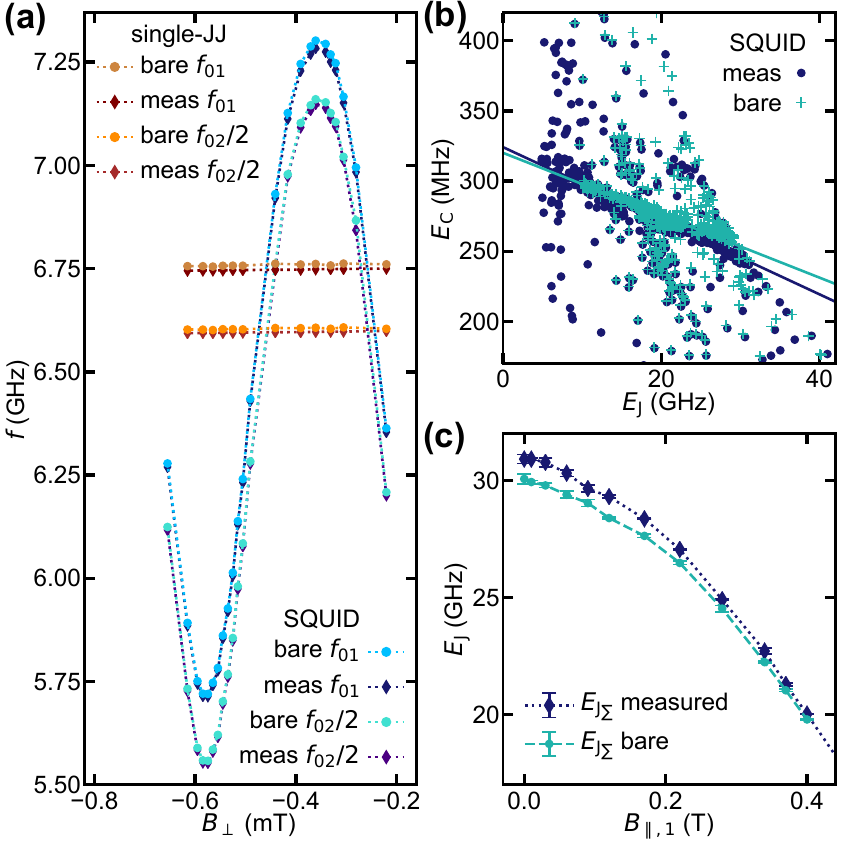}
   \end{center}
   \caption{Estimating bare transmon parameters.
      \textbf{(a)} Fitting a two-qutrit-one-cavity Hamiltonian to the measured frequencies, we can estimate the bare transmon frequencies.
      Closer to the cavity resonance frequency ($f_\mathrm{c}=8.1$~GHz), the hybridization is stronger leading to a larger correction.
      \textbf{(b)} The $\EC$-$\EJ$ correlation, as described in \cref{fig:EC_vs_EJ} gives a slightly altered linear trend for $\EC$s from bare frequencies.
      \textbf{(c)} Refitting the estimated bare SQUID flux arches we obtain a downward correction of $\EJ$ by \SI{3}{\percent}. With increasing field the correction becomes smaller, as the qubit frequencies and consequently the hybridization with the cavity mode decrease.
   }
   \label{fig:Hamiltonian_fit}
\end{figure}

Close to the cavity frequency, the anharmonicity of the transmon is modified by hybridization with the cavity.
To estimate this effect and correct for it, we fit a two-qutrit-one-cavity Hamiltonian of the form 
\begin{equation}
H=H_0+H_\mathrm{coupling}+H_{qq}.
\label{eq:two_qutrit_one_cavity_Hamiltonian}
\end{equation}
Here, $H_0$ is the uncoupled Hamiltonian for two qutrits and a resonator,
\begin{equation}
\begin{split}
   H_0 = & \; \hbar\omega_\mathrm{c} a^\dagger a  \\
        &   + \omega_{01, 1} \ket{1}_1 \bra{1}_1 + \omega_{02, 1} \ket{2}_1 \bra{2}_1  \\
        &   + \omega_{01, 2} \ket{1}_2 \bra{1}_2 + \omega_{02, 2} \ket{2}_2 \bra{2}_2,
\end{split}
\end{equation}
with $\omega_\mathrm{c}=2\pi\fcav$ the cavity angular frequency and creation/annihilation operators $\omega_\mathrm{c}$, $a$ and $a^{\dagger}$.
$H_\mathrm{coupling}$ then models the qutrit-cavity interaction in the rotating-wave approximation, but avoiding the dispersive approximation:
\begin{equation}
\begin{split}
   H_\mathrm{coupling}= \; & \hbar g_1 \left[ \left ( \ket{0}_1 \bra{1}_1 + \sqrt{2} \ket{1}_1 \bra{2}_1 \right ) a+ \mathrm{c.c.} \right] \\
             +  & \hbar g_2 \left[ \left ( \ket{0}_2 \bra{1}_2 + \sqrt{2} \ket{1}_2 \bra{2}_2 \right ) a+ \mathrm{c.c.} \right] .
\end{split}
\end{equation}
Here, $g_{1,2}$ denote the coupling strength between the respective qutrit and the cavity.
Finally, $H_{qq}$ would be a direct qubit-qubit interaction.
However, we measured the qubit-qubit avoided crossings at several fields and can bound the interaction to below \SI{1}{\mega\hertz}.
For the fit, we only used data away from the avoided crossing and neglected this term.
Having a data set of dressed transitions $\omega_\mathrm{c}$, $\omega_{01}^{(q1)}$, $\nicefrac{\omega_{02}^{(q1)}}{2}$, $\omega_{01}^{(q2)}$ and $\nicefrac{\omega_{02}^{(q2)}}{2}$ we fit the respective energy levels of \cref{eq:two_qutrit_one_cavity_Hamiltonian} to these transitions (see \cref{fig:Hamiltonian_fit} \textbf{(a)}). 
Approximating the bare cavity frequency by its high power limit, $\fcav=\SI{8.107}{\giga\hertz}$, we obtain the cavity-qutrit couplings and bare qutrit frequencies.
The couplings $g_{1,2}$ show a slight frequency dependence, which is expected as the transmon dipole moment is dependent on $\EJ$ and $\EC$~\cite{Koch07}.
In our fit range, we can approximate $\nicefrac{g_{1,2}}{2\pi}=\SI{57}{\mega\hertz}+0.01\fzeroone$, meaning $\nicefrac{g_{1,2}}{2\pi}$ ranging from \SIrange{100}{130}{\mega\hertz}.
Refitting the estimated bare SQUID frequency dependence with $\Bperp$, we obtain more accurate values for the transmon $\EJ$ and $\EC$.
The bare and dressed values for $\EC$ and $\EJ$ are compared in \cref{fig:Hamiltonian_fit} \textbf{(b)} and \textbf{(c)}.
The downward correction of $\EJ$ is less than \SI{3}{\percent}.
With increasing field, the correction becomes even smaller, as the qubit frequencies and consequently the hybridization with the cavity mode decrease.
The $\EJ$ presented in the main text is based on the bare levels when the transmon frequencies are close to the cavity.

\begin{figure}
  \begin{center}
  \includegraphics[width=\columnwidth]{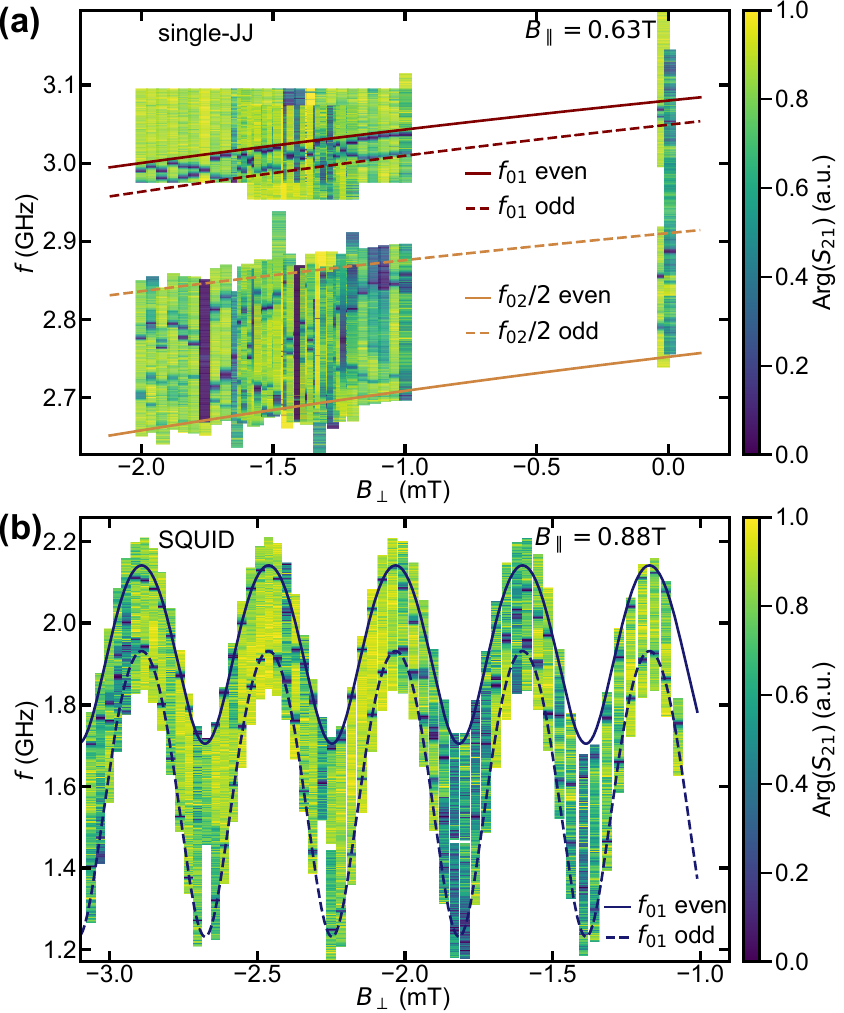}
  \end{center}
  \caption{
  Example data for extracting $\EJ$ and $\EC$ from charge-parity splitting for the single-JJ (panel \textbf{(a)}) and SQUID transmon (panel \textbf{(b)}).
  Only the maximum splitting between the peaks needs to be estimated as any value in between can be observed for different $n_g$ in the Hamiltonian.
  The changes in $n_g$ that modulate the splitting are happening at a timescale slower than the measurements, therefore one can observe the opening and closing of the charge-parity. 
  For the single-JJ example, we could also observe $\fzerotwoovertwo$, for the SQUID example we were in a regime where we could no longer observe $\fzerotwoovertwo$ and had to rely solely on $\fzeroone$. 
  }
  \label{fig:charge_parity_splitting}
\end{figure}

For the highest fields, charge-parity splitting becomes a dominant effect in the transmon spectrum, as the $\nicefrac{\EJ}{\EC}$ ratio becomes small.
In spectroscopy, we observe peaks for the odd and even parity subspace and the charge offset $n_g$ randomly changes.
Example data sets and fits for single-JJ and SQUID qubit are shown in \cref{fig:charge_parity_splitting} \textbf{(a)} and \textbf{(b)}.
The Hamiltonian remains the same as \cref{eq:transmon_hamiltonian}, but we evaluate it for $n_g=0$ and $n_g=0.5$ to have the two parity branches. 
Populations of those two states as well as the exact value of $n_g$ are random and drift.
In order to extract $\EJ$ and $\EC$ in this regime we fitted transitions from the charge-parity split Hamiltonian to bound the experimental data. 

\section{Sensitivity analysis of noise in $\Bperp$}
\label{sec:apdx_sensitivity_analysis}

\begin{figure}
  \begin{center}
    \includegraphics[width=\columnwidth]{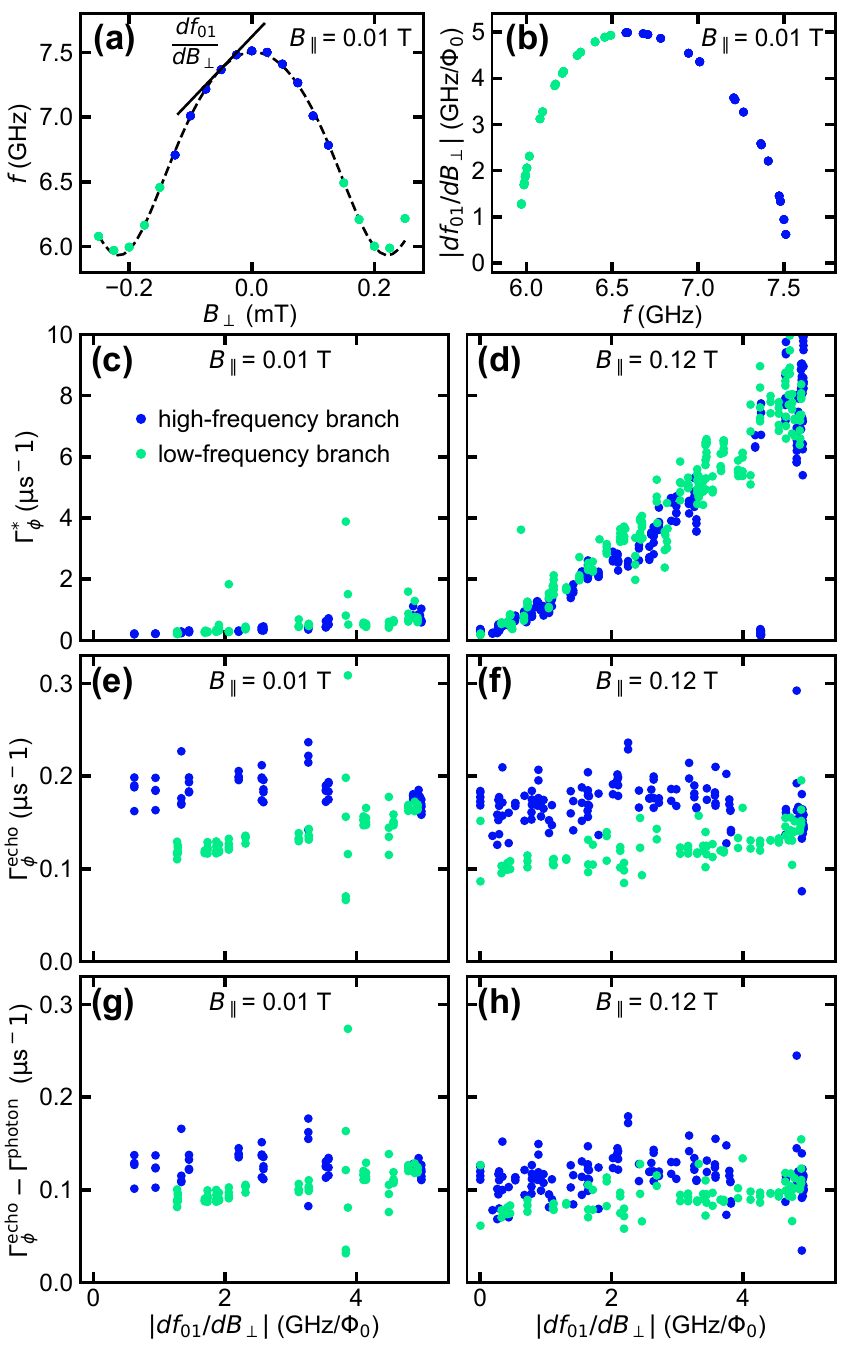}
  \end{center}
  \caption{
    \textbf{(a)} $f_{01}$ of the SQUID qubit vs. the out-of-plane field $\Bperp$ for an example flux oscillation of the SQUID at in plane field $\Bparone=0.01\,\mathrm{T}$. The tangent line indicates the sensitivity at that point.
    \textbf{(b)} out-of-plane field sensitivity $\sens$ as a function of frequency for the complete in plane field dataset.
    Data in all figures are color coded for the low-frequency and high-frequency sensitivity branch.
    \textbf{(c)} and \textbf{(d)} The pure Ramsey dephasing rates of the SQUID qubit as a function of $\sens$ at $\Bparone=\SI{0.01}{\tesla}$ and $\Bparone=\SI{0.12}{\tesla}$. 
    \textbf{(e)} and \textbf{(f)} The pure echo dephasing rates of the SQUID qubit as a function of $\sens$ at $\Bparone=\SI{0.01}{\tesla}$ and $\Bparone=\SI{0.12}{\tesla}$.
    \textbf{(g)} and \textbf{(h)} The pure echo dephasing rates of the SQUID qubit subtracting our estimate of photon shot noise as a function of $\sens$ at $\Bparone=\SI{0.01}{\tesla}$ and $\Bparone=\SI{0.12}{\tesla}$. 
  }

  \label{fig:pure_echo_dephasing}
\end{figure}

Any noisy parameter that tunes the transmon frequency reduces its coherence.
The SQUID transmon frequency and therefore its coherence are sensitive to noise in the perpendicular magnetic field component $\Bperp$.
This noise can be on-chip flux noise or setup-related, e.g. noise in the current source powering the magnet coils or vibrations of the sample with respect to the vector magnet.
The sensitivity $\sens$ determines the extent to which noise in $\Bperp$ reduces the transmon coherence.
To calculate it, we fit the the flux dependence of the SQUID frequency using \cref{eq:SQUID_EJ_vs_phi}.
For every frequency, we calculate $\sens$ from the fitted curve (\cref{fig:pure_echo_dephasing} \textbf{(a)}). 
We can then plot the $\sens$ as a function of SQUID transmon frequency $\fzeroone$ (\cref{fig:pure_echo_dephasing} \textbf{(b)}).
The main parameters that contribute to the $\sens$ as a function of frequency are the SQUID period $\Bphinaughtsquid$, and top and bottom sweetspot frequencies (as well as $\EC$ to a lesser degree).
The sensitivity is given in units of GHz/$\Phi_{0}$, because our fitted curve also contains the periodicity in $\Bperp$, thus we rescale the x-axis in units of $\Phi_{0}$.

To quantify the transmon coherence, we calculate the pure dephasing time
\begin{equation}
  \frac{1}{\Ttwo} = \Gphi+ \frac{1}{2\Tone}
\end{equation}
to separate the contribution of the dephasing rate and the lifetime.
Having measured $\Tone$, $\Ttwostar$ and $\Techo$ as a function of $\Bperp$ over at least one period of the SQUID, we can plot $\Gphi$ against $\sens$.
\cref{fig:pure_echo_dephasing} \textbf{(c)} and \textbf{(d)} illustrate the case for $\Gstar$, and \cref{fig:pure_echo_dephasing}  \textbf{(e)} and \textbf{(f)} for $\Gecho$.
We observe a linear dependence on sensitivity
\begin{equation}
  \label{eq:Gphi_sensitivity}
  \Gphi = a \left \vert \frac{d\fzeroone}{d\Bperp}\right\vert+b,
\end{equation}
where $a$ describes slow noise that scales with $\Bperp$ and the offset $b$ accounts for flux-independent noise contributions.
As illustrated in the four examples \cref{fig:pure_echo_dephasing} \textbf{(c)}-\textbf{(f)} we generally find such a linear trend for the Ramsey data, but not for the echo experiments where we can see a clear difference between the high-frequency and low-frequency branch of the sensitivity.
Therefore, it is only for $\Gstar$ that we can extract the noise parameters $a$ and $b$ for each in-plane magnetic field and analyze them as a function of $\Bparone$.
The result for $a$ are shown and discussed in \cref{sec:qubit_coherence_sensitivity_analysis}.

Compared to the Ramsey experiments, $\Gecho$ should be more robust against low-frequency noise and give insights into faster noise like on-chip flux noise as opposed to slow setup-related vibrational noise.
However, as shown in \cref{sec:timedomain_limits} the echo dephasing rates are partially limited by photon shot noise in the cavity which is strongly frequency dependent.
Assuming this photon-shot-noise contribution, we subtract it as a frequency-dependent background.
With the background subtraction, the gap between the upper and lower sensitivity branch (which differ in frequency) is reduced, as seen in \cref{fig:pure_echo_dephasing} \textbf{(g)} and \textbf{(h)}.
For the Ramsey data, the two branches of the sensitivity are consistent with the same linear trend to begin with, because the noise in $\Bperp$ is more strongly limiting $\Gstar$.
However, the photon-shot-noise subtracted $\Gecho$ becomes relatively flat, so there is likely another $\Bperp$-independent noise source limiting $\Gecho$.
There is also no strong change in $\Gecho$ as a function of $\Bparone$, until we reach fields where we also become more charge sensitive.
Thus our data cannot give information on the characteristic $1/f$-like flux noise ubiquitous in cQED~\cite{Bylander11,Kumar16}.
If this noise originates from local paramagnetic fluctuators, it is expected to depend on applied magnetic fields~\cite{LuthiPhD19}, therefore repeating this experiment with improved coherence times could give insights into this.

\section{Additional relevant limits on coherence}
\label{sec:timedomain_limits}

\begin{figure}
   \begin{center}
      \includegraphics[width=\columnwidth]{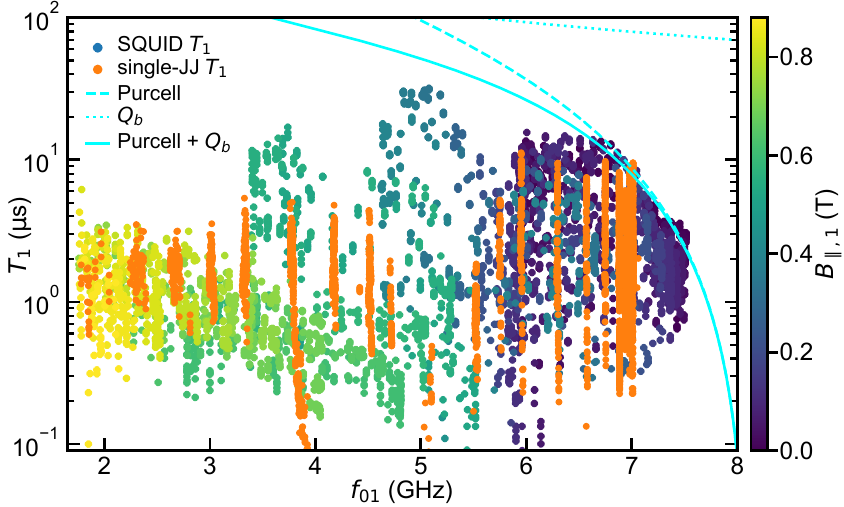}
   \end{center}
   \caption{
      $\Tone$ vs transmon $\fzeroone$ for every in-plane magnetic field for both the single-JJ and the SQUID transmon.
      Purcell decay to the cavity mode is limiting both qubits at high frequencies.
      For similar frequencies but different $\Bparone$, single-JJ and SQUID transmon show very different features, e.g. a dip and a peak around \SI{5.0}{\giga\hertz}.
      The observed lifetimes are not following a frequency dependent loss model, as in~\cite{Luthi18}, except for $\fzeroone\geq$\SI{5.8}{\giga\hertz}.
   }
   \label{fig:t1_vs_frequency}
\end{figure}

\begin{figure}
  \begin{center}
    \includegraphics[width=\columnwidth]{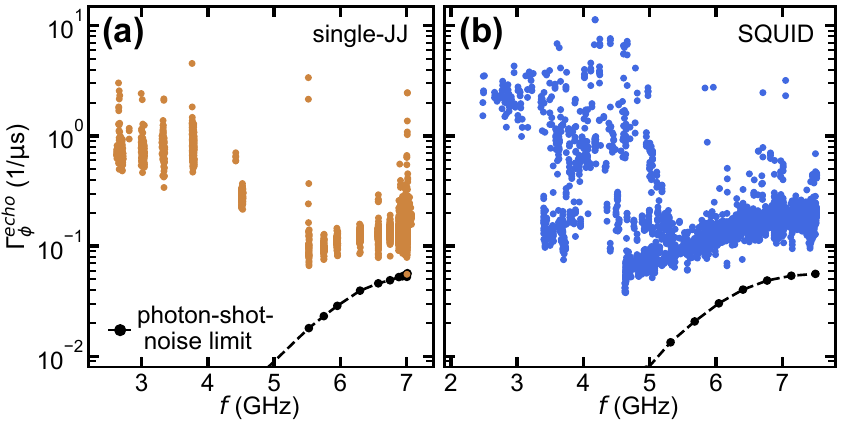}
  \end{center}
  \caption{
    \textbf{(a)} The pure echo dephasing rate of the single JJ transmon against the transmon transition frequency.
    The black circles model photon shot noise with $T_\mathrm{cav}=\SI{76}{\milli\kelvin}$.
    \textbf{(b)} Similar to \textbf{(a)}, but for the SQUID transmon.
  }
  \label{fig:pure_echo_dephasing_vs_freq}
\end{figure}

Here, we will present our understanding of additional relevant limits on transmon relaxation time and coherence in this work.
While the qubits are not reaching the current state of the art in coherence times, this is likely largely due to insufficient shielding and filtering (see Ref.~\onlinecite{SOM}).
We will discuss Purcell and quasiparticle limits on $\Tone$ and show evidence that $\Techo$ is limited by photon shot noise.
In the main text above, we already discussed mechanical vibrations as a likely culprit limiting $\Ttwostar$ at high $\Bparone$.
Vortex loss is discussed in \cref{sec:vortex_loss}.
We do not discuss the limits imposed on qubit dephasing by charge noise and quasiparticle dynamics in more detail in this work, but they might be different from transmons with thicker aluminum films.

\subsection{Limits to $\Tone$}

An important factor to consider in the qubit energy relaxation is frequency. 
\cref{fig:t1_vs_frequency} shows all measured $\Tone$ values as a function of frequency for both qubits.
There are several frequency-dependent mechanisms limiting $\Tone$.
The Purcell effect imposes a limit based on the coupling to the readout cavity $\Tone^{\mathrm{Purcell}}=\nicefrac{\delta^2}{(g^2 \kappa)}$, where $\delta=\fzeroone-\fcav$ is the detuning between qubit and cavity mode, $g$ is the qubit-cavity coupling, and $\kappa$ the cavity linewidth.
Instead of $\kappa$, the quality factor $Q_\mathrm{tot} = \nicefrac{2\pi\fcav}{\kappa}$ is often quoted.
We find $Q_\mathrm{tot}\approx5800$ at low fields but at $\Bparone\ge\SI{0.5}{\tesla}$ we find $Q_\mathrm{tot}\approx3800$.   
Surprisingly, our measured $\fcav$ and $Q_\mathrm{tot}$ together with our estimates for $g$ give a Purcell limit that some of our measured $\Tone$ values exceed.
Over a larger range, transmons limited by dielectric loss often exhibit an overall trend in $\Tone$ that roughly follows the form $\Tone^\mathrm{b}=\nicefrac{Q_\mathrm{b}}{2\pi\fzeroone}$, where $Q_\mathrm{b}$ is a background quality factor~\cite{Wang15}.
We cannot observe such a trend convincingly but did include it in \cref{fig:t1_vs_frequency} as a guide to the eye, setting $Q_\mathrm{b}=3.5\times10^6$.

As the magnetic field suppresses the superconducting gap, it is important to consider quasiparticle-induced relaxation.
To estimate this effect, the superconducting gap needs to be estimated in absolute terms.
As discussed in \cref{sec:ip_magnetic_field_dependence} we estimate the in-plane critical field to be  $\Bcritpar=\SI{1.03}{\tesla}$.
Assuming a Ginzburg-Landau closing of the superconducting gap (\cref{eq:GL_Gap}), we find the gap to be reduced by only $\SI{\sim50}{\percent}$ at $\Bparone=\SI{0.88}{\tesla}$.
The data taken during the cooldown suggests $T_\mathrm{crit}\approx\SI{1.2}{\kelvin}$ which we can use to estimate the gap at zero magnetic field via $\Delta_0=1.764\,k_\mathrm{B} T_\mathrm{crit}$~\cite{Bardeen57,Tinkham04}.
With these values we can calculate an estimate for the quasiparticle-induced relaxation rate~\cite{Catelani11}
\begin{equation}
  \Gqp=2\frac{8\EJ\EC\xqp}{\fzeroone}\sqrt{\frac{2\Delta}{h\fzeroone}},
\end{equation}
where
\begin{equation}
  \xqp=\sqrt{2\pi\frac{k_\mathrm{B} T_\mathrm{R}}{\Delta}}\exp\left({-\frac{\Delta}{k_\mathrm{B} T_\mathrm{R}}}\right)
\end{equation}
is the normalized quasiparticle density assuming a thermal equilibrium temperature $T_\mathrm{R}$. 
Here, $h$ is the Planck constant and $k_\mathrm{B}$ is the Boltzmann constant.
The remaining free parameter is the quasiparticle bath temperature $T_\mathrm{R}$.
At \SI{0.88}{\tesla} we measure a $\Tone$ of \SI{2.4}{\micro\second} for the SQUID transmon, which is at a frequency of \SI{1.8}{\giga\hertz}.
With these values we can roughly bound $T_\mathrm{R}$ to be $\le\SI{90}{\milli\kelvin}$.
More importantly, we do not observe any sharp decrease in $\Tone$ with $\Bparone$ that would signal loss mechanism becoming dominant~\cite{Luthi18}.
This suggests that up to the highest field we measured, $\Tone$ is not significantly limited by quasiparticles, yet.

\subsection{Limits to $\Ttwo$}

Turning to echo coherence times $\Techo$, we will now estimate the limit imposed by photon shot noise in the cavity.
As can be seen in \cref{fig:fig4}, $\Gecho$ slightly decreases with $\Bparone$ for $\Bparone<\SI{400}{\milli\tesla}$.
Looking at $\Gecho$ as a function of qubit frequency (\cref{fig:pure_echo_dephasing_vs_freq}) is more revealing in this context.
A dependence on qubit frequency is expected using a model for photon shot noise~\cite{Rigetti12,Clerk07}.
In the dispersive limit, the qubit-cavity interaction is reduced to a term of the form $\chi a^{\dagger}a \sigma_\mathrm{z}$.
Accordingly, the qubit frequency depends on the cavity photon number $a^{\dagger}a$ via the dispersive shift $\chi$, which for a transmon is given by
\begin{equation}
  \begin{split}
    \chi = & g^2 \alpha_\mathrm{tr}\left[\frac{1}{\delta (\delta +\alpha_\mathrm{tr})}\right.  \\
    &  \left.-\frac{1}{(\delta - 2\fzeroone)(\delta-\alpha_\mathrm{tr}-2\fzeroone)}\right],  
  \end{split}
\end{equation}
where $\alpha_\mathrm{tr}=f_{12}-f_{01}$ is the negative transmon anharmonicity.
Thus, thermal fluctuations in the cavity photon number lead to a dephasing rate 
\begin{equation}
  \Gamma^\mathrm{photon} = \frac{\kappa}{2} \mathrm{Re}\left[\sqrt{\left(1+\frac{2i\chi}{\kappa}\right)^2 +\left(\frac{8i\chi n_\mathrm{th}}{\kappa}\right)}-1\right].
\end{equation}
The thermal cavity photon number $n_\mathrm{th} = \left(\exp(\nicefrac{h f_\mathrm{c}}{k_\mathrm{B}T_\mathrm{cav}})-1\right)^{-1}$ is given by Bose-Einstein statistics. 
The only free parameter is the cavity temperature $T_\mathrm{cav}$.
As the photon shot noise limit on $\Techo$ mainly depends on the transmon frequency, we show $\Gecho$ versus $\fzeroone$ for both transmons in \cref{fig:pure_echo_dephasing_vs_freq}.
Using $T_\mathrm{cav}=\SI{76}{\milli\kelvin}$ we approximately reproduce the smallest $\Gecho$ we measured for the single-JJ transmon.
The data point is admittedly an outlier, but both the underlying $\Tone$ and $\Techo$ measurements have good signal-to-noise ratio and fits, however, the $\Tone$ could have fluctuated.
In that case the photon shot noise would likely be more severe.
Even for this low estimate, at qubit frequencies above \SI{5}{\giga\hertz}, photon shot noise is a significant contribution to $\Techo$.

\clearpage

\renewcommand{\theequation}{S\arabic{equation}}
\renewcommand{\thefigure}{S\arabic{figure}}
\renewcommand{\thetable}{S\arabic{table}}
\renewcommand{\thesection}{S\arabic{section}}

\setcounter{section}{0}    
\setcounter{equation}{0}
\setcounter{figure}{0}
\setcounter{table}{0}
\setcounter{page}{1}

\onecolumngrid
\begin{center}
  \textbf{\large Supplementary Material for ``Magnetic-field resilience of 3D transmons with thin-film Al/AlO$_x$/Al Josephson junctions approaching 1 T''}\\[.3cm]
  J. Krause,$^{1,*}$ C. Dickel,$^{1,*}$  E. Vaal,$^{1,2}$  M. Vielmetter,$^{1}$  J. Feng,$^{1}$ \\[.1cm] 
  R. Bounds,$^{1}$ G. Catelani,$^{2}$ J. M. Fink,$^{3}$ Yoichi Ando,$^{1}$\\[.2cm]
  {\itshape 
  ${}^1$ Physics Institute II, University of Cologne, Zülpicher Str. 77, 50937 Köln, Germany \\[.1cm] 
  ${}^2$ JARA Institute for Quantum Information (PGI-11),\\ Forschungszentrum Jülich, 52425 Jülich, Germany \\[.1cm]
  ${}^3$} Institute of Science and Technology Austria, Klosterneuburg, Austria \\[.1cm]
  ${}^*$ These authors contributed equally to this work.\\
(Dated: \today)
\end{center}
\twocolumngrid

This supplement provides experimental details and additional data supporting the claims in the main text.

\renewcommand{\appendixname}{}

\section*{Experimental setup}
\label{sec:setup}

\begin{figure}
  \begin{center}
  \includegraphics[width=\columnwidth]{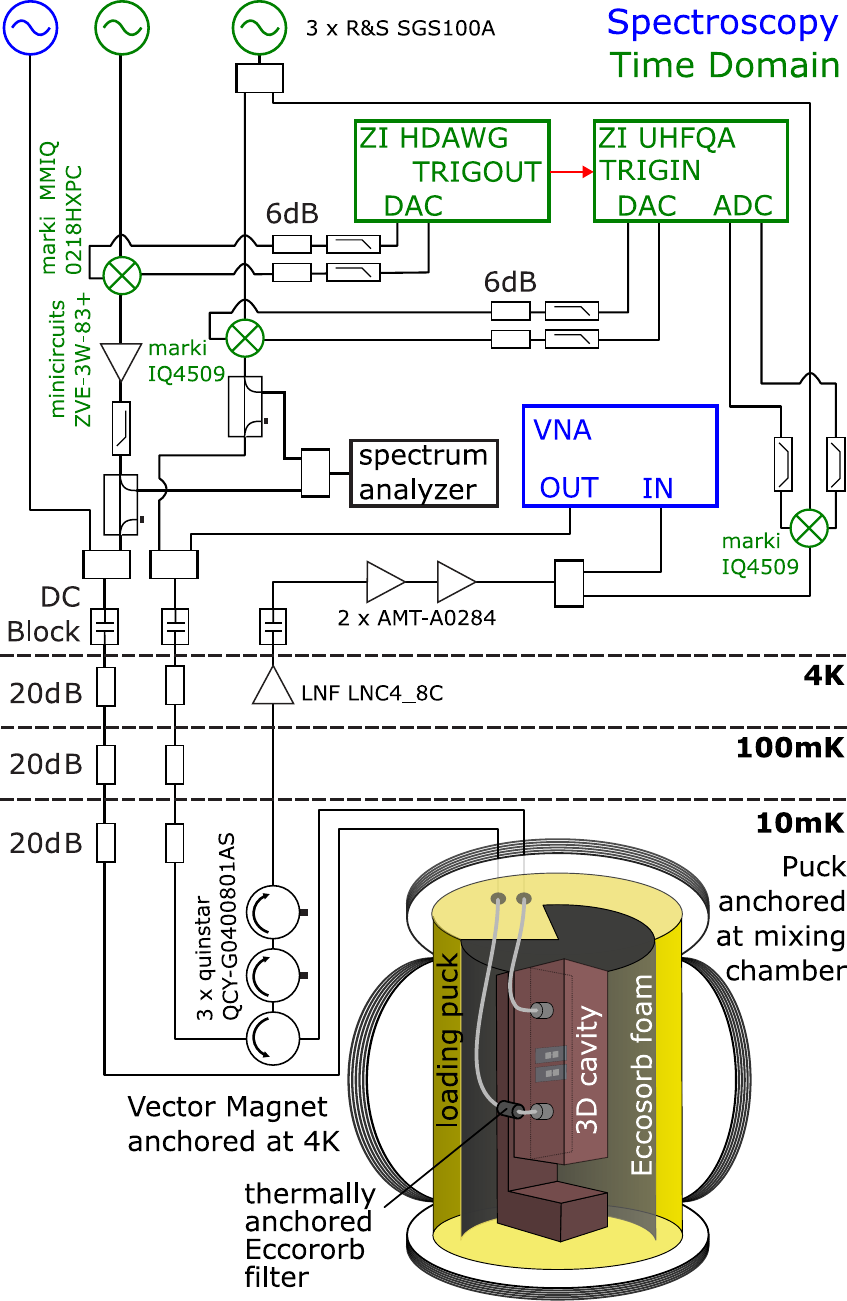}
  \end{center}
  \caption{Wiring diagram of the experiment with the setup for qubit spectroscopy using a two-port VNA and an additional microwave source and the time-domain setup using ADCs, DACs and Mixers.
           }
  \label{fig:wirind_diagram}
\end{figure}

The transmons are mounted in a 3D cavity which is loaded into a bottom-loading dilution refrigerator (Triton 500, Oxford Instruments) with a nominal base temperature of $\sim$\SI{10}{\milli\kelvin}. 
In our bottom-loading dilution refrigerator, the radiation shielding setup is likely not optimal and could partially account for the somewhat limited $\Tone$ relaxation times and high cavity temperature. 
With the strong magnetic fields, usual magnetic shields used in cQED experiments consisting of $\mathrm{\mu}$-metal and superconducting boxes around the samples could not be used.
We use Eccosorb LS-26 foam  inside the loading puck to improve radiation shielding (see \cref{fig:wirind_diagram}).
In a similar setup, radiation as a limiting factor was suspected~\cite{KrollPhD19}.
A detailed wiring diagram of the experiment can be found in \cref{fig:wirind_diagram}.
The fridge wiring is loosely based on \cite{Krinner19}.
While the maximum $\Tone$ reported here is $\sim$\SI{30}{\micro\second}, we previously measured \SI{60}{\micro\second} on a different single-JJ transmon with a frequency of $\sim$\SI{4}{\giga\hertz} in a nominally identical dilution refrigerator, therefore we do not believe our setup to impose a general strong $\Tone$ limit.

A 3-axis vector magnet is used to apply magnetic fields to the sample.
In this paper we do not give data for field cooled transmon, but we note that when we performed an initial cooldown with the high-field current supply of our magnet connected to the out-of-plane axis this already lead to frequency jumps in the transmons. 
Therefore, during cooldowns the magnet power supply was always turned off and the magnet was grounded. 
Everything was turned on once base temperature was reached.
Not all magnet axes were connected at all times in order to minimize noise.
For the in-plane axis we mainly used the Oxford Instruments Mercury iPS power supply, which can supply sufficient current to reach a magnetic field of \SI{1}{\tesla} .
We also used a Keithley 2461 as a current source to compare the noise level.
The Keithley 2461 can only reach magnetic fields of \SI{0.16}{\tesla}.
For the out-of-plane axis, we used a Keithley 2450 for the data presented in the main text which can only reach \SI{0.016}{\tesla}.
For the data presented in \cref{fig:large_boop_cavity}, we used the Keithley 2461 for the out-of-plane magnetic field coil.

The 3D cavity is machined from oxygen free copper. 
It has two symmetrically coupled ports, making it possible to measure it both in reflection and transmission. 
We mostly used reflection measurements as they allow an estimation of the intrinsic quality factor.

\section*{Large out-of-plane magnetic fields and hysteresis}
\label{sec:Large_Boop}

\begin{figure}
  \begin{center}
  \includegraphics[width=\columnwidth]{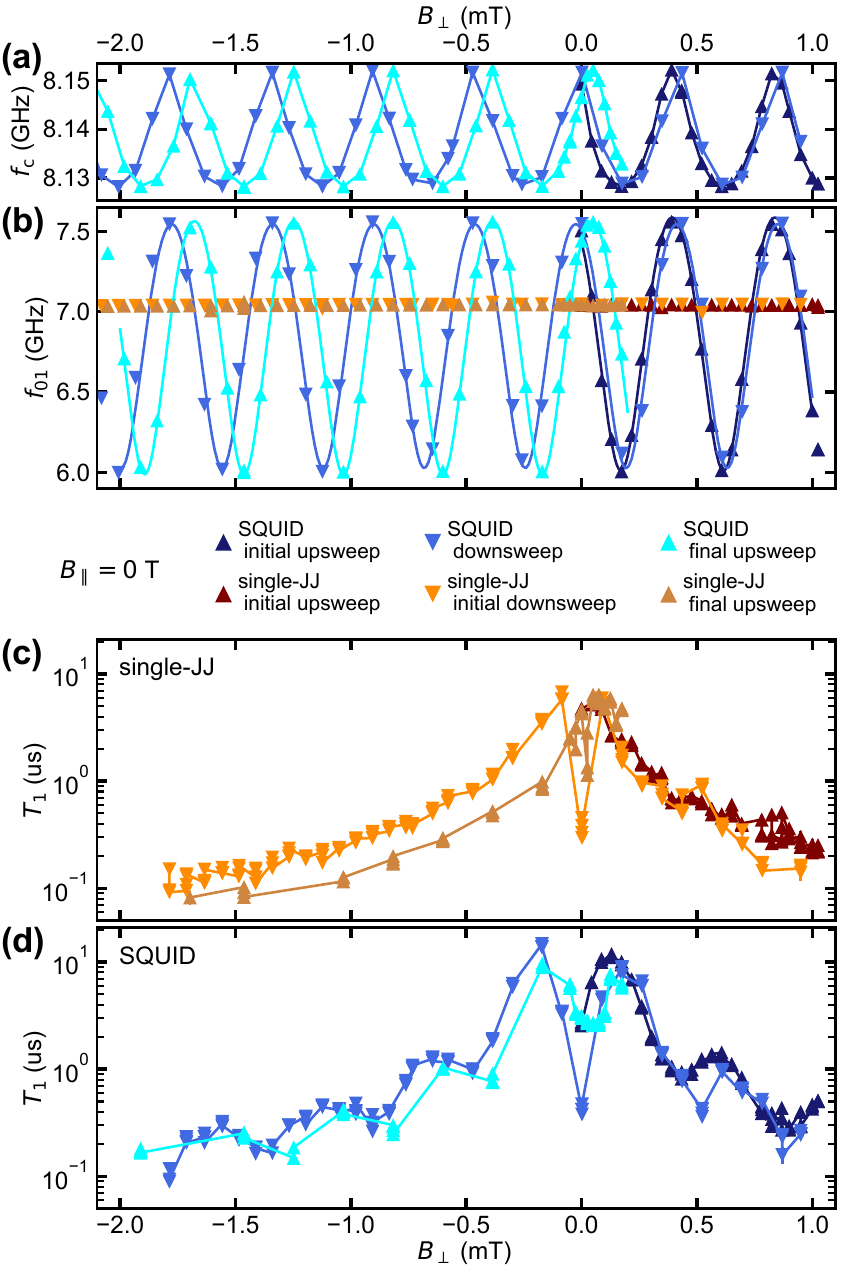}
  \end{center}
  \caption{
         \textbf{(a)} and \textbf{(b)} Cavity frequency $f_\mathrm{c}$ and the qubit frequency $\fzeroone$ for both transmons as a function of $\Bperp$ for $\Bpar=0$.
         The Data were taken by scanning up from $\Bperp = \; $\SI{0}{\milli\tesla}, then down to outside the plot range and back up to $\Bperp = $\SI{0.2}{\milli\tesla}, with different markers and colors for the segments.
         $f_\mathrm{c}$ clearly follows the SQUID transmon oscillations. 
         For the initial upsweep and downsweep the hysteresis is visible but small, but upon the return from larger negative $\Bperp$ hysteresis is a considerable fraction of a flux quantum.
         \textbf{(c)} and \textbf{(d)} $\Tone$ for the single-JJ and SQUID transmon as a function of $\Bperp$.
         The SQUID data shows oscillations that are due to the frequency dependence of $\Tone$, generally showing longer $\Tone$ at the bottom sweetspot.
         Strangely, $\Tone$ drops for both qubits on the way back to $\Bperp=$\SI{0}{\milli\tesla}.
         These data were taken in an initial cooldown in another nominally identical dilution refrigerator, thus qubit frequencies slightly vary compared to the data reported in the main text. 
           }
  \label{fig:large_boop_data}
\end{figure}

In this section, we discuss the large $\Bperp$ regime looking at the qubit coherence and the persistence of SQUID oscillations as well as hysteresis in $\Bperp$ as a factor complicating experiments.
\cref{fig:large_boop_data} shows data on cavity frequency $f_\mathrm{c}$ and transmon frequency $\fzeroone$ as well as $\Tone$ for both transmons as a function of $\Bperp$.
Data taking started immediately after a thermal cycle and only the magnet coil corresponding to $\Bperp$ was connected.
This data was obtained in a separate cooldown in a different nominally identical dilution refrigerator. 
The data is color coded for the different scan segments (\SI{0}{\milli\tesla} to \SI{1}{\milli\tesla}, \SI{1}{\milli\tesla} to \SI{-10}{\milli\tesla} and \SI{-10}{\milli\tesla} to \SI{0.2}{\milli\tesla}).
Time domain measurements were stopped around \SI{-2}{\milli\tesla}, so for the large field excursion we do not plot the entire curve.
For the initial two scans, only extending to a small range of $\Bperp$, the change in field direction leaves visible but small hysteresis.
However, after the excursion to large $\Bperp$ on the downward scan, the upward scan coming back shows a hysteresis that is a significant part of a flux quantum.
Notably, in the scanned range here, we do not observe flux jumps. 
It is important to mention in passing, that in \cref{fig:large_boop_data} where qubit spectroscopy data is plotted as well as cavity measurements, it is obvious that the cavity oscillations follow the SQUID transmon oscillations which repels the cavity.
Later on, we will look at oscillations in the cavity frequency and assume that they are due to the SQUID oscillation in the qubit.

\begin{figure}
  \begin{center}
  \includegraphics[width=\columnwidth]{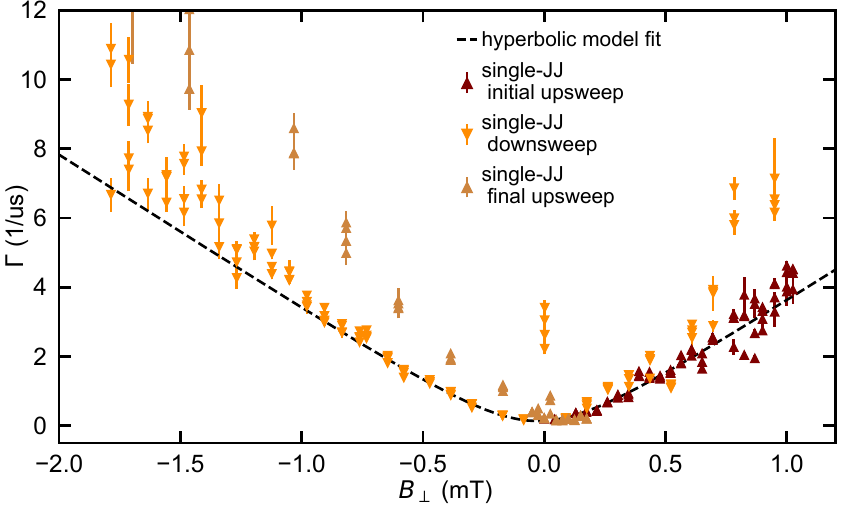}
  \end{center}
  \caption{
         Relaxation rate $\Gamma$ as a function of $\Bperp$.
         The data are the same as in \cref{fig:large_boop_data} \textbf{(c)}. 
         The dashed line shows a fit of the hyperbolic model (see main text). 
         It is obvious that hysteresis is an important factor here, thus we fit the model through the downsweep data that was taken without changing sweep direction.
         }
  \label{fig:large_boop_gamma}
\end{figure}

For this data we can also look at $\Gamma = 1/\Tone$ as a function of $\Bperp$ for the single-JJ transmon (see \cref{fig:large_boop_gamma}). 
This larger range in $\Bperp$ shows more clearly that an expected linear dependence of $\Gamma$ on $\Bperp$ can only be seen for larger $\Bperp$.
We can only measure $\Gamma$ precisely over roughly two orders of magnitude here, for small $\Tone$, the error bars on $\Gamma$ become large.

\begin{figure}
  \begin{center}
  \includegraphics[width=\columnwidth]{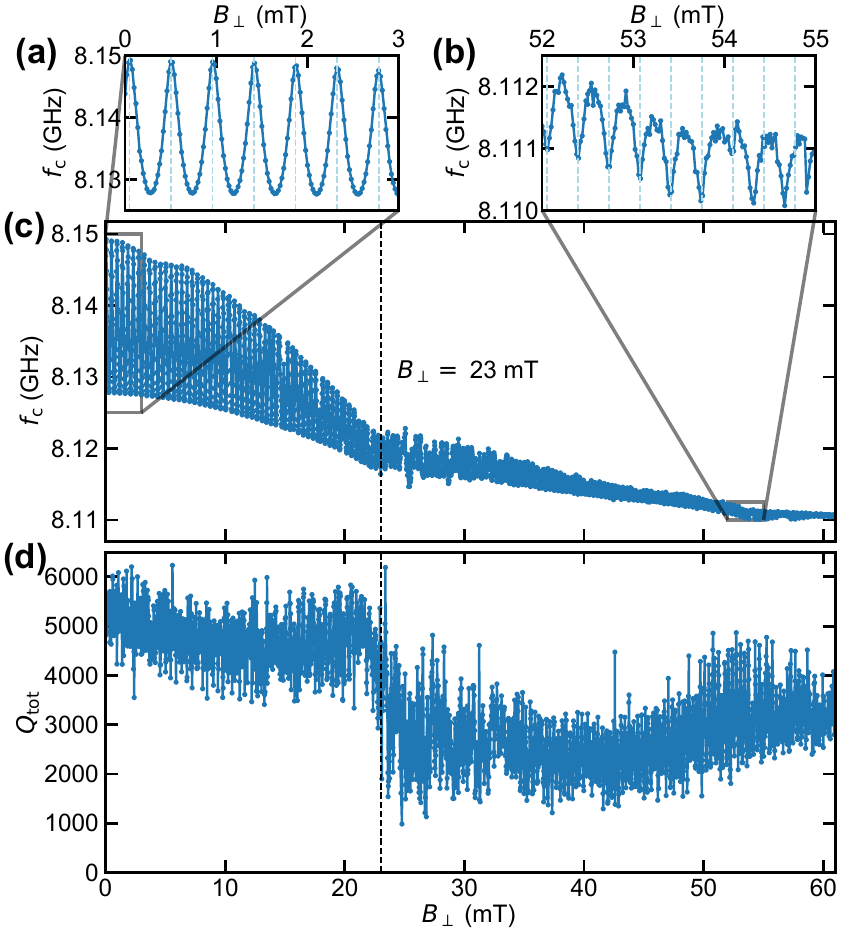}
  \end{center}
  \caption{\textbf{(c)} Cavity frequency as a function of $\Bperp$.
           Zoom-in panels (\textbf{(a)} and \textbf{(b)}) show the low-field and high-field regime.
           While there is clearly a change in the oscillations around $\Bperp\approx$ \SI{23}{\milli\tesla}, oscillations persist in our measurable range.
           However, the period decreases from $\sim$\SI{0.455}{\milli\tesla} to $\sim$\SI{0.34}{\milli\tesla} (dashed vertical lines).
           \textbf{(d)} Cavity total quality factor $Q_\mathrm{tot}$ as a function of $\Bperp$.
           $Q_\mathrm{tot}$ is reduced from $\sim$5000 to $\sim$2500 around $\Bperp\approx$\SI{23}{\milli\tesla}, also suggesting a change in the system at that point.
           Interestingly, towards large $\Bperp$, $Q_\mathrm{tot}$ seems to recover.
           The $Q_\mathrm{tot}$ data is noisy here because the cavity reflection measurements were done with a large bandwidth to save time.
           }
  \label{fig:large_boop_cavity}
\end{figure}

In order to estimate the out-of-plane critical field, we measured the dependence of the cavity frequency on $\Bperp$ expecting a breakdown of SQUID oscillations (\cref{fig:large_boop_cavity}).
There is a clear change in the SQUID oscillation pattern around \SI{23}{\milli\tesla}.
At this point, the cavity quality factor $Q_\mathrm{tot}$ also decreases.
The most likely explanation for this decrease is the transition from superconducting to normal state of the capacitor pads: from $B_{c2} \equiv \Phi_0/(2\pi \xi^2) = \SI{23}{\milli\tesla}$, we find $\xi \approx \SI{120}{\nano\meter}$, in good agreement with the rough estimate made in Appendix D of the main text.

Unexpectedly, the oscillations in \cref{fig:large_boop_cavity} remain visible across the range that we measured, suggesting that superconductivity might persist up to $\Bperp=\SI{60}{\milli\tesla}$.
The period of the oscillations becomes smaller suggesting the effective area of the SQUID loop increases, which one would expect, as more field penetrates the superconductor.
Flux jumps become more frequent at higher fields.
It is plausible that the critical fields for the large capacitor pads is different from that of the thin leads and JJ region.
However, our estimate for the perpendicular critical field based on a Ginzburg-Landau theory fit to the magnetic-field dependence of the single-JJ transmon frequency yields $\Bcritperp\approx\SI{33}{\milli\tesla}$ (presented in the main text).
While spectroscopy was only performed up to $\Bperp\sim\SI{8}{\milli\tesla}$ and Ginzburg Landau theory is approximate in this regime, we do not believe that superconductivity can still be present at the highest fields reached in our measurements.
A possible explanation for the observed oscillations could be the Altshuler-Aronov-Spivak effect~\cite{Altshuler81}. 
This effect should be relevant in rings with circumference longer than the mean free path $\ell$ and at most of the order of the phase coherence lenght $L_\phi$. 
In our device, the circumference is about \SI{8}{\micro\meter} and $\ell \sim \SI{10}{\nano\meter}$ (see main text). 
Estimates for $L_\phi$ in aluminum at temperatures around $T_\mathrm{c}$ and zero magnetic field are of the order of 1-\SI{2}{\micro\meter}~\cite{Gijs84,Chandrasekhar85}. 
On one hand, it is not unreasonable to expect $L_\phi$ to be longer well below $T_\mathrm{c}$ (cf.~\cite{Pierre03}); on the other hand, $L_\phi$ should be suppressed by the applied field to a fraction of a micron. 
Although superconducting fluctuations could in part compensate for such a suppression (see also~\cite{Shimshoni93}), confirmation or refutation of our hypothesis is left to future work.

\begin{figure}
  \begin{center}
  \includegraphics[width=\columnwidth]{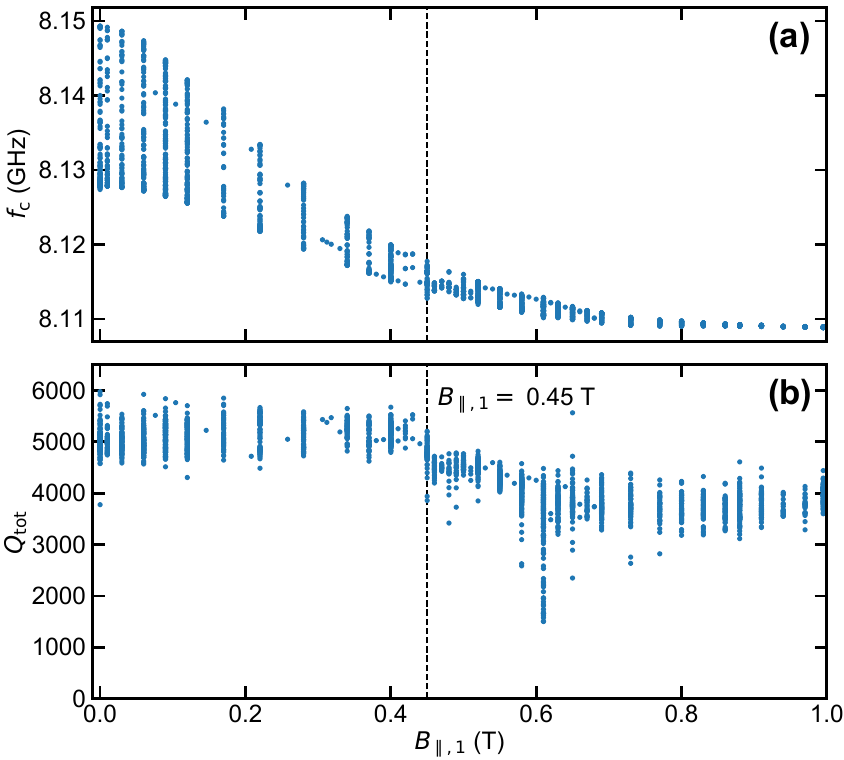}
  \end{center}
  \caption{\textbf{(a)} Cavity frequency as a function of $\Bparone$.
           We show the cavity measurements taken for the qubit spectroscopy and time domain measurements of the main text but reject some outliers as it is a large dataset.
           At each $\Bparone$ there is a range of $\fcav$ for different $\Bperp$.
           \textbf{(b)} Cavity total quality factor $Q_\mathrm{tot}$ as a function of $\Bparone$.
           We see an almost step-like drop in $Q_\mathrm{tot}$ around $\Bparone\approx\SI{0.45}{\tesla}$.       
           }
  \label{fig:cavity_vs_Bparone}
\end{figure}

It is interesting to compare the $\fcav$ and $Q_\mathrm{tot}$ as a function of $\Bperp$ with the cavity data as a function of $\Bparone$ (\cref{fig:cavity_vs_Bparone}). 
These cavity measurements were taken during the qubit spectroscopy and time domain measurements presented in the main text.
At each $\Bparone$ a range of measurements with different $\Bperp$ were taken, therefore there is usually a range of different $\fcav$ and $Q_\mathrm{tot}$.
The data looks qualitatively remarkably similar to \cref{fig:large_boop_cavity}.  
While the reduction in $\fcav$ with $\Bparone$ is just due to the reduced frequencies of the transmons presented in the main text, the drop in the maximum $Q_\mathrm{tot}$ around $\Bparone\approx\SI{0.45}{\tesla}$ is unexpected.
It is less severe than the drop in $Q_\mathrm{tot}$ as a function of $\Bperp$, but contrary to the data in \cref{fig:large_boop_cavity}, we do not reach the estimate for $\Bcritparone$.
From the other measurements presented in the main text, we can also exclude that this is an effect of misalignment of the magnet axes to the sample.

\section*{SQUID oscillation stability vs $\Bparone$ and $\Bpartwo$}
\label{sec:SQUID_oscillation_stability}

One of the more puzzling observations presented here is the SQUID oscillation instability.
SQUID oscillations are usually a robust feature that would only disappear if superconductivity is fully suppressed in the circuit or in one or both of the constituting JJs.
While we can see clear SQUID oscillations for $\Bparone$ in the range from \SI{0}{\tesla}-\SI{1}{\tesla}, there is an instability region roughly between \SI{0.4}{\tesla} and \SI{0.5}{\tesla}.
The step-like drop in the $Q_\mathrm{tot}$ of the cavity (\cref{fig:cavity_vs_Bparone}) occurs at $\Bparone\approx\SI{0.45}{\tesla}$ in the instability region.
In this region we also observed that the cavity did not remain stable in time with jumps in its resonance frequency occurring every few minutes.
Therefore we did not perform detailed time-domain measurements in this region and did not have good measurements of the SQUID transmon.
Spectroscopy measurements of the single-JJ transmon (where the frequency could be roughly predicted) were possible in between cavity jumps.
We speculate the instability of the cavity frequency is due to the SQUID transmon being unstable in frequency, but we cannot exclude other options such as direct coupling to spin baths conclusively.
Interestingly, there are outliers in our data where the cavity is at higher frequencies than would be consistent with the range of dispersive shifts for the expected maximum SQUID frequency.

The spin hypothesis arose because for a g-factor of $\sim$2, spin precession frequencies could become resonant with the qubits or cavity in the magnetic field range we explored.
However, we do see the cavity instability and disappearance of SQUID oscillations at different fields for $\Bpartwo$ (\cref{fig:SQUID_oscillation_stability_Bpar2}).
\cref{fig:SQUID_oscillation_stability_Bpar2} \textbf{(a)} shows back and forth scans for small $\Bpartwo$ in the -\SI{40}{\milli\tesla} to \SI{40}{\milli\tesla} region.
Already at fields with $\left \vert {\Bpartwo} \right \vert >$\SI{20}{\milli\tesla}, the regular SQUID oscillations disappear.
It is for this reason, that we have not studied the qubit spectroscopy and coherence as a function of $\Bpartwo$. 
There seems to be a bistability in the SQUID oscillation offset at low fields, possibly due to residual ferromagnetism in our setup that can be sensed by the SQUID.
Our cavity connectors are unfortunately nickel coated, so they could be responsible. 
The chaotic regions at larger fields are clearly visible and not reproducible.
Interestingly, at higher fields, there are regions where regular SQUID oscillations are again visible in $\Bpartwo$ (\cref{fig:SQUID_oscillation_stability_Bpar2} \textbf{(b)}).

The anisotropy for different in-plane axes weakens the case for spins and we believe it points towards the spurious junctions in the leads as culprits, since the spurious JJ geometry is asymmetric for different in-plane directions.
The spurious JJs in our devices are not as well defined as the ones discussed in Ref.~\onlinecite{Schneider19}, as in our case the entire leads as well as the capacitor pads are formed by the two aluminum layers with the oxide in between.
Given that the leads from the capacitor pads to the JJ are essentially a spurious JJ, it makes sense to consider the Fraunhofer pattern for this region.
The lead width was measured in SEM pictures to be \SI{410}{\nano\meter} and the SQUID arms are of similar size. 
The field to thread a flux quantum through the spurious JJ in the leads and SQUID arms in the $\Bparone$ direction is on the order of \SI{0.45}{\tesla}, if we assume an insulator thickness similar to the one of the JJs discussed in the main text.
This seems to coincide with the instability region in $\Bparone$ between \SI{0.4}{\tesla} and \SI{0.5}{\tesla}.
In the Fraunhofer lobe region, the supercurrent is suppressed and the devices become very field sensitive, which could also to some extent explain the unstable behavior.
The expected $\EJ$ for the spurious JJs is very large, such that away from the fields where supercurrent is suppressed, the effects could be negligible. 
It seems to contradict this theory that the single-JJ transmon, which has identical leads to the SQUID transmon, can often be found close to where it would be expected in spectroscopy in the instability region.
But the strong anisotropy would be consistent with the asymmetric shape of the spurious JJs in the $\Bparone$ and $\Bpartwo$ directions.
Looking at the device from the $\Bpartwo$ direction the length of the spurious JJ is huge (the leads are \SI{100}{\micro\meter} long) and thus much lower fields could cause Fraunhofer lobes.
In the large capacitor pads and maybe even the leads, pinholes likely limit the overall area that could be considered as one continuous JJ.
The onset at \SI{20}{\milli\tesla} in the $\Bpartwo$ direction would correspond to a length scale of \SI{10}{\micro\meter}.
It is puzzling that stable SQUID oscillations are observed again from $\Bpartwo>\SI{0.6}{\tesla}$ to the maximum measured field of \SI{0.9}{\tesla}.
Given the confusing and unstable data, we did not investigate this further.

\begin{figure}
  \begin{center}
  \includegraphics[width=\columnwidth]{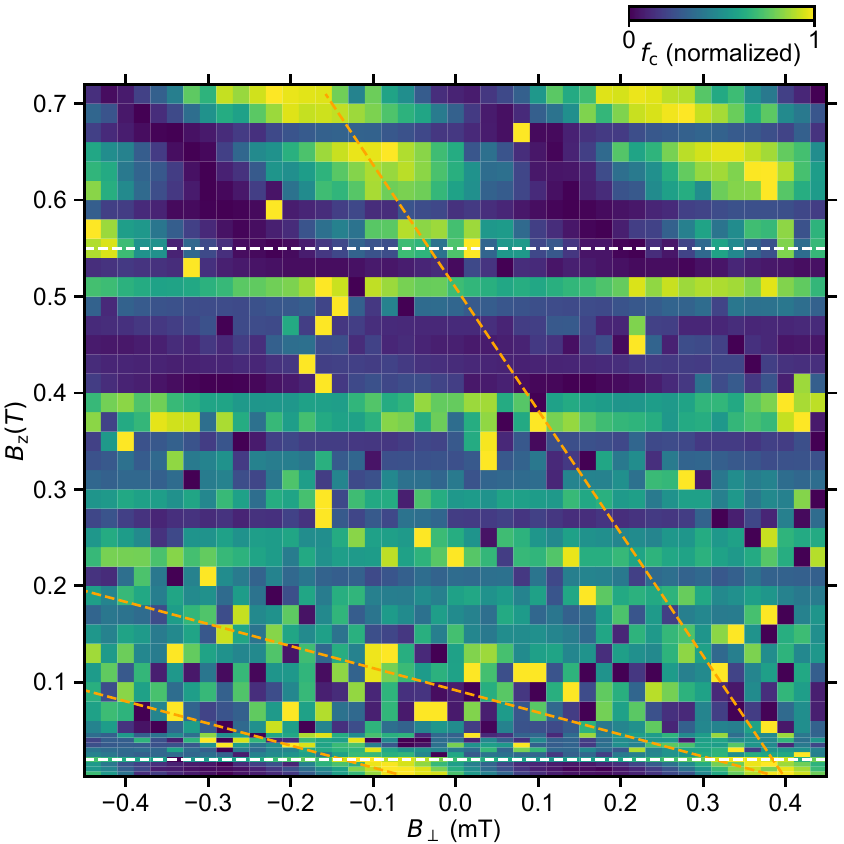}
  \end{center}
  \caption{
    Cavity frequency as a function of $\Bperp$ and $\Bz$, which is largely the $\Bpartwo$ direction.
    The data are normalized line-by-line to highlight the periodicity in $\Bperp$ due to the SQUID oscillations.
    Clear SQUID oscillations can only be made out up to $\Bz=\SI{0.02}{\tesla}$ (bottom white dashed line).
    At high fields they recommence clearly around \SI{0.55}{\tesla} (upper white dashed line).
    The linear dependence of the offset of the oscillation as a function of $\Bz$ is slightly different for the low-field and high-field regime (orange dashed lines).
    }
  \label{fig:SQUID_oscillation_stability_Bz}
\end{figure}

\begin{figure}
  \begin{center}
  \includegraphics[width=\columnwidth]{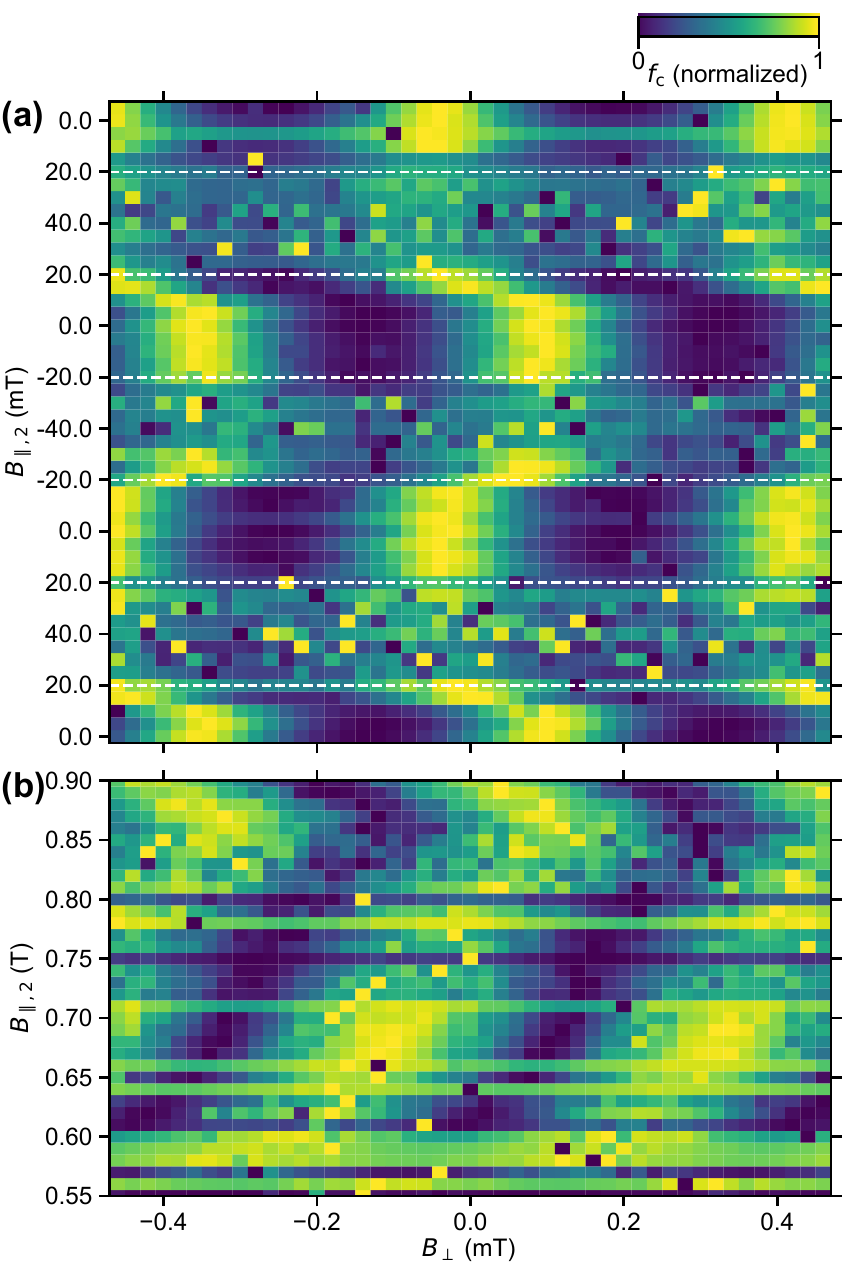}
  \end{center}
  \caption{
    \textbf{(a)} Cavity frequency as a function of $\Bperp$ and $\Bpartwo$.
    The data are normalized line-by-line.
    These data were taken in a low $\Bpartwo$ regime scanning from \SI{0}{\milli\tesla} to +\SI{40}{\milli\tesla} to -\SI{40}{\milli\tesla} to \SI{40}{\milli\tesla} and back to \SI{0}{\milli\tesla}.
    In the region from \SI{20}{\milli\tesla} to \SI{40}{\milli\tesla} (and \SI{-20}{\milli\tesla} to \SI{-40}{\milli\tesla}) oscillations disappear and the cavity frequency as a function of the magnetic fields looks random.
    Dashed lines that mark the boundaries of the stable oscillation regime are added for clarity.
    The alignment looks good as the SQUID offset is essentially stable as $\Bpartwo$ is changed.
    \textbf{(b)} Cavity frequency (line-by-line normalized for contrast) as a function of $\Bperp$ and $\Bpartwo$ for large $\Bpartwo$.
    Cavity oscillations return around 600~mT and persist up to 900~mT where we stopped measuring in this dataset.
    The period remains similar at high $\Bpartwo$.
    }
  \label{fig:SQUID_oscillation_stability_Bpar2}
\end{figure}

\clearpage

\end{document}